\documentclass[pre,superscriptaddress,twocolumn,a4paper,showpacs]{revtex4-1}
\usepackage{graphics}
\usepackage{amssymb}
\usepackage{wasysym}
\usepackage{latexsym}
\usepackage{graphicx}
\usepackage{epsfig}
\usepackage{subfigure}
\begin{document}

\title{Comfort-driven mobility  produces spatial fragmentation  in Axelrod's model}  

\author{Sandro M. Reia}
\affiliation{Instituto de F\'{\i}sica de S\~ao Carlos,
  Universidade de S\~ao Paulo,
  Caixa Postal 369, 13560-970 S\~ao Carlos, S\~ao Paulo, Brazil}

\author{Paulo F. Gomes}
\affiliation{Instituto de F\'{\i}sica de S\~ao Carlos,
  Universidade de S\~ao Paulo,
  Caixa Postal 369, 13560-970 S\~ao Carlos, S\~ao Paulo, Brazil}
\affiliation{Instituto de Ci\^encias Exatas e Tecnol\'ogicas, Universidade Federal de Goi\'as, 
75801-615  Jata\'{\i}, Goi\'as, Brazil }
  
  \author{Jos\'e F.  Fontanari}
\affiliation{Instituto de F\'{\i}sica de S\~ao Carlos,
  Universidade de S\~ao Paulo,
  Caixa Postal 369, 13560-970 S\~ao Carlos, S\~ao Paulo, Brazil}

\begin{abstract}
Axelrod's model for the dissemination of culture combines  two key ingredients of social dynamics: social influence, through which people become more similar when they interact, and homophily, which is the tendency of individuals to interact preferentially with similar others. In Axelrod's model,  the agents are fixed to the nodes of a network and  are allowed to interact with a predetermined set of peers only, resulting in  the frustration of a large number of agents that end up  culturally isolated. Here we modify  this  model by allowing  the agents  to move away from their cultural opposites and  stay put when near their cultural likes. The comfort, i.e., the tendency of an agent to stay put  in a neighborhood, is determined by the cultural similarity with its neighbors. The less the comfort, the higher the odds that the agents will move apart a fixed step size. 
 We find that the comfort-driven mobility  fragments severely  the influence network for low  initial cultural diversity, resulting in a  network  composed of only  microscopic components  in the thermodynamic limit. For high initial cultural diversity and intermediate values of the step size, we find that a macroscopic component coexists with the microscopic ones.  The transition between these two fragmentation regimes changes from continuous to discontinuous as the step size increases. In addition, we find that for both very small  and very large  step sizes the influence network is severely fragmented. 
%
%\keywords{Communication networks \and Cooperative processes \and  Imitative learning \and NK landscapes}
\end{abstract}

\maketitle

\section{Introduction}\label{sec:intro}
Homophily (i.e., the tendency of individuals to interact preferentially with similar others) and social influence (i.e., 
the tendency of individuals to become  more similar with whom they interact) have long  been  perceived as major factors that influence
social phenomena like segregation, inter-group bias and inequality, to mention only a few 
\cite{Lazarsfeld_48,Castellano_09,Galam_12}. The understanding of the ways these factors impact social organization  has been  considerably expanded by the study of  the agent-based model  proposed by the political scientist Robert Axelrod in the late 1990s \cite{Axelrod_97}. Axelrod's model  offered  a simple  quantitative approach to address the dissemination of culture among interacting agents in a society.

In Axelrod's   model,   the agents  are represented by  strings  of
cultural features of length $F$, where each feature can take on a certain number $q > 1$ of distinct states
(i.e., $q$ is  the common number of states that each feature can assume).  Hence the parameter $F$ represents the  complexity of the society, since the different features are associated to different individual characteristics that are subject to social influence such as  language, education, class, politics,  religion, etc.,
whereas the number of states per feature $q$ represents the (potential)  cultural diversity of the society -- the larger $q$, the greater the number of options for the
cultural features \cite{Axelrod_97}. 

The homophily factor is  accounted for
by  the assumption that the interaction between  two agents takes place
with probability proportional to their cultural similarity  (i.e., proportional to the number of states they have in common), whereas social
influence  enters Axelrod's  model by forcing the agents to become more similar when they interact. Thus, there is a  positive feedback loop between homophily and social influence: similarity leads to interaction, and interaction leads to still more similarity. Somewhat surprisingly,
in spite of this homogenizing mechanism,  Axelrod's model exhibits global polarization (i.e., a stable multicultural regime) for large $q$ in the case the agents are
fixed to  the sites of a square lattice and interact with their nearest neighbors only \cite{Axelrod_97}. Variations of the original model have revealed  that the relaxation of the homophilic interaction rules and the expansion of the interaction neighborhoods  favor  cultural homogenization (i.e., a stable monocultural regime)  \cite{Klemm_03a,Klemm_03b,Klemm_03c,JEDC_05,Reia_16}.  

 In the context of social organization,  an important issue is the coevolution of the  cultural states  of the agents and the structure of the interaction or influence network
(i.e., who interacts with whom) on the same time scale \cite{Zimmerman_04,Pacheco_06,Vazquez_07,Vazquez_08,Kimura_08,Herrera_11}. In fact, whereas in the original Axelrod model the agents  are forced to interact with a predetermined group of agents (usually their nearest neighbors), we  expect that in a more realistic scenario  the agents would actively seek their likes and avoid their opposites.  Such a scenario  was considered for  the  nonlinear voter model \cite{Min_17}, the weighted social network model \cite{Murase_19} and the Sznajd model \cite{Benatti_19}, where  the network topology is allowed to change  by (probabilistically) rewiring the links between agents with a bias towards the creation of links connecting agents with similar opinions or cultures.  

In this paper we take a different approach and allow  the agents  to move in a square box by performing steps of fixed size $\delta$ in random directions in the plane. The agents are initially located at the nodes of random geometric graphs (RGGs) \cite{Gilbert_61},
where each node is randomly assigned geometric coordinates and  two nodes are connected if the Euclidean distance between them is smaller than or equal to a certain threshold $d$. The set of agents  connected to a particular agent  defines its influence neighborhood. The RGGs were  introduced  in the 1960s to model wireless communication networks \cite{Gilbert_61} and its connectivity properties were  subsequently investigated for threshold phenomena \cite{Penrose_03,Dall_02}. More recently, RGGs have been used to model synchronization \cite{Guilera_09}, opinion dynamics \cite{Zhang_14}, epidemic spreading \cite{Estrada_16} and epistemic communities \cite{Reia_19}  in scenarios where the agents are geographically constrained in  certain regions. 

The comfort-driven mobility is introduced in the model  by assuming that the probability that a given agent stays put  is proportional to the maximum value of its cultural similarity evaluated over all agents within its  influence neighborhood.  In addition, the agent moves with certainty if it is isolated, i.e., if 
 its influence neighborhood is empty.  Hence, an agent feels more comfortable (in the sense that it is less likely to move) in a location where at least one of its neighbors has a high homophily with it. These  rules of motion  are akin to those used in the  modeling of the  dynamics of human interactions with the cultural similarity playing  the role  of the social appeal between the individuals \cite{Zhao_11,Starnini_13}. 
We find that, for large systems, endowing the agents of Axelrod's model with the capacity to move in a plane following the comfort-driven  rules of motion
 fragments the  initially connected RGG into a macroscopic number of components and greatly suppress the culturally isolated agents, so
 most agents are comfortable in the absorbing configurations.

The rest of this paper is organized as follows. In section \ref{sec:model} we describe
 the  rules of motion in the two-dimensional physical space where the agents roam freely as well as  the  rules that govern  their interactions  and determine how their cultural states change in time.
 In  section \ref{sec:res} we present and analyze the results of our simulations, emphasizing the influence of the step size  on the  spatial and cultural organization of Axelrod's model with mobile agents.  Finally, section \ref{sec:disc} is reserved to our concluding remarks.

\section{Model}\label{sec:model}

\begin{figure}  
\centering
  \includegraphics[width=0.48\textwidth]{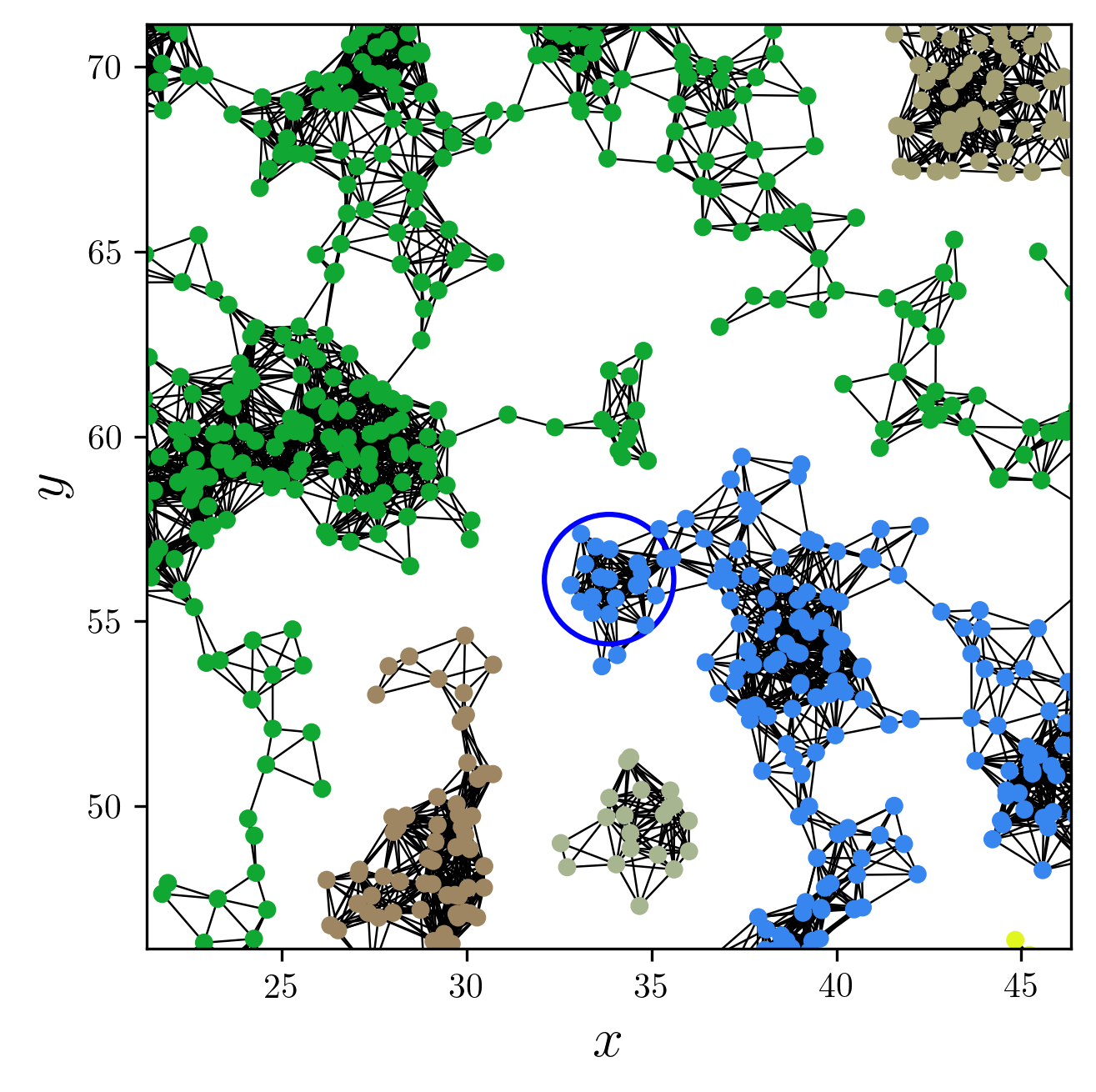}
\caption{Snapshot of a small portion of an absorbing configuration with $N=2^{15}$ agents and initial cultural diversity $q = 10$. The display of the full configuration requires that the x and y axes  range from 0 to   $2^{15/2} \approx 181$.  Agents within a distance $d = \alpha $,  with  $\alpha=1.75$,  are connected by a link.  The circle of radius $d$ centered at the central agent determines its influence neighborhood. The comfort-driven mobility with step size $\delta=9$ causes the fragmentation of the initial  random geometric graph into a large number of components.
 }  
\label{fig:1}  
\end{figure}

We consider a system of $N$ agents placed  in a   square box of linear size $L$ with periodic boundary conditions.  In the initial configuration, the coordinates $x$ and $y$ of each agent are chosen  randomly and uniformly over the  length $L$. The density of agents
$\rho = N/L^2$,  which we fix to $\rho = 1$ throughout this paper, yields the spatial scale for the interaction range $d$  and the step size  $\delta$. In fact, since the effective  area of an agent is $1/\rho$, the  quantity $d_0 = 1/\sqrt{\rho} = 1$ will be the standard  to measure all distances in our study.    More pointedly, we measure the distance $d$ within which interactions between agents are allowed in units of $d_0$, i.e., $d = d_0 \alpha=\alpha$ with $\alpha > 0$. The set of agents inside a circle of radius $d$ centered at a particular agent constitutes the influence neighborhood from where that agent selects a peer to interact with. Linking any  two agents at a distance smaller  than $d$   produces an undirected graph that we refer to as the influence network (see Fig.\ \ref{fig:1}).  Thus, as already pointed out,  the initial disposition of the agents corresponds to the classic random geometric graph  \cite{Gilbert_61}. 
We note that the fixed value of  the density $\rho$ is inconsequential, provided we use $d_0$ as the standard for measuring distances in the square box. However, the choice $\rho=1$ is consistent with the density of the regular square lattice, which was  used in most  studies of the Axelrod model \cite{Axelrod_97}.

%-----------------------------------------------------
%-----------------------------------------------------

In Axelrod's model, the initial states of the $F$  cultural  features of the  agents are  drawn randomly from a uniform distribution on the integers $1, 2, \ldots, q$. Once the initial configuration (i.e., the  positions of the agents in the square box and their cultural features)  is set, the dynamics  proceeds as follows. It begins with the selection of an agent at random, the so-called target agent,  
and  comprises two stages. The first stage is the motion on the square box and the second stage is the social interaction. To decide whether to move or not, the  target agent evaluates its cultural similarity with all agents in its influence neighborhood and singles out  the maximum value, which we denote by  $a_m \in \left [0,1\right]$. (We recall that the cultural similarity between two agents is simply the fraction of features they have in common \cite{Axelrod_97}.) Here we assume that the target agent moves with probability $1- a_m$  and  stays put with probability $a_m$. In the case the  influence neighborhood of the target agent is empty, it moves with probability $1$. These  rules of motion imply   a sort of `repulsion to the different'  that prompts an agent  to move away more frequently from neighborhoods that lack  its cultural likes.  

In the case the  target agent decides to move, 
 an angle $\theta \in \left [ 0, 2 \pi \right )$  is chosen randomly to give the direction of motion   and then a fixed step of length $\delta \geq 0$  is taken on that direction, similarly to the procedure used  in the  modeling of the  dynamics of human interactions \cite{Starnini_13} as well as in the  study of the effects of random motility on cooperative problem-solving systems  \cite{Freitas_19}.
  Once the target agent is at the new position, a circle of radius $d = \alpha$  is drawn around it so that its (new) influence neighborhood is determined.  Then the  social  interaction  stage  sets in:  an agent within the  influence neighborhood of the target agent is chosen at random and they interact with probability equal to  their cultural similarity.   An interaction consists of selecting at random one of the distinct features and making the
selected feature of the target agent equal  to  the corresponding  feature of its randomly chosen peer \cite{Axelrod_97}. 
In the case the  target agent stays put,  only the social interaction stage  is implemented.

This procedure is repeated until  the dynamics enters an absorbing configuration.
According to the social interaction rule, absorbing configurations are such that agents within the influence neighborhood of a target agent are either identical to  or completely different from it with respect to their cultural features.  We note that the agents are not necessarily static in the absorbing configurations:  in principle, an  (non-interacting) agent  that shares no cultural feature  with any other agent in the system will keep moving forever without affecting or being affected by  the established stationary social organization.
However, we can easily identify this situation, which  happens  very rarely for $\delta > 0$  since by construction the comfort-driven mobility aims at preventing the appearance of  uncomfortable, i.e., culturally isolated, agents.

Once the dynamics reaches an absorbing configuration, we count the number of  cultural domains ($\mathcal{N}_d$) and record the size of the largest one ($\mathcal{S}_d$), as usual
\cite{Klemm_03a,Klemm_03b,Klemm_03c,JEDC_05,Reia_16}. In time, a cultural domain is defined  as  a connected subgraph where the agents have the same culture (i.e., they share all cultural features). Hence, cultural diasporas are considered  different cultural domains. In addition and more importantly, because of the rules of motion of the agents that, in principle, could allow them to organize themselves in isolated clusters or components, we  measure also the number of  components ($\mathcal{N}_c$) and 
the size of the largest component ($\mathcal{S}_c$) of the influence network. Since a component can sustain many cultural domains, we have  $\mathcal{N}_d \geq \mathcal{N}_c$ and  $\mathcal{S}_c \geq \mathcal{S}_d$. As all these quantities are bounded by the number of agents $N$, in section \ref{sec:res} we will characterize the absorbing configurations in terms of the densities   $n_d = \mathcal{N}_d/N$,  $s_d = \mathcal{S}_d/N$, $n_c = \mathcal{N}_c/N$ and  $s_c = \mathcal{S}_c/N$.

Moreover, since our goal is to study   the  comfort-driven mobility, our study will focus mainly  on the influence of the step size or mobility parameter $\delta$  on the statistical properties of the absorbing configurations. 
Accordingly, we will fix the parameter that  determines the radius of the influence neighborhoods to  $\alpha = 1.75$. For large $N$ (and $\rho=1$) this  choice produces initial configurations that are  random geometric graphs with  average degree per agent  $\langle k_i \rangle \approx 9.62$, so the initial influence networks are almost surely connected graphs \cite{Penrose_03,Dall_02}. In addition,  we will fix the number of cultural features to $F=3$ since this is the 
minimum value of $F$ for which the ordered and disordered regimes are stable in a large range of values of $q$ in the static case. Use of larger values of $F$ makes the convergence to the absorbing configurations prohibitively slow for large $N$. In the brief study of the static limit $\delta =0$, we will consider also the case $F=2$ in order to highlight the distinct  nature of the phase transitions for $F=2$ and $F=3$.

 For the sake of illustration, we show in Fig.\ \ref{fig:1}  a snapshot of  a portion of an absorbing configuration  of  a system of $N=2^{15}$ agents with $q=10$ and step size $\delta =9$. The influence neighborhood of the central agent is shown as a  circle in the figure. The different components of this small portion  of  the influence network can be easily identified   due to the absence of links between them. Next we will quantify the puzzling effects of the mobility parameter $\delta$ on the connectedness  of the influence networks.

\section{Results}\label{sec:res}

The measures  we use to characterize the statistical properties of the absorbing configurations represent
averages  over (typically) $10^3$ independent runs, which differ initially by  the cultural states of the agents as well as by  their positions  on the square box. As mentioned before,  in the study  of the effects of the mobility parameter $\delta > 0$ we fix the radius of the influence neighborhood $\alpha$ and the cultural complexity $F$ of the system to  $\alpha =1.75$ and $F=3$. The case $F=2$ is considered only for
the static agents  scenario, $\delta =0$.

\subsection{Static agents}

\begin{figure} 
\centering  
\includegraphics[width=0.48\textwidth]{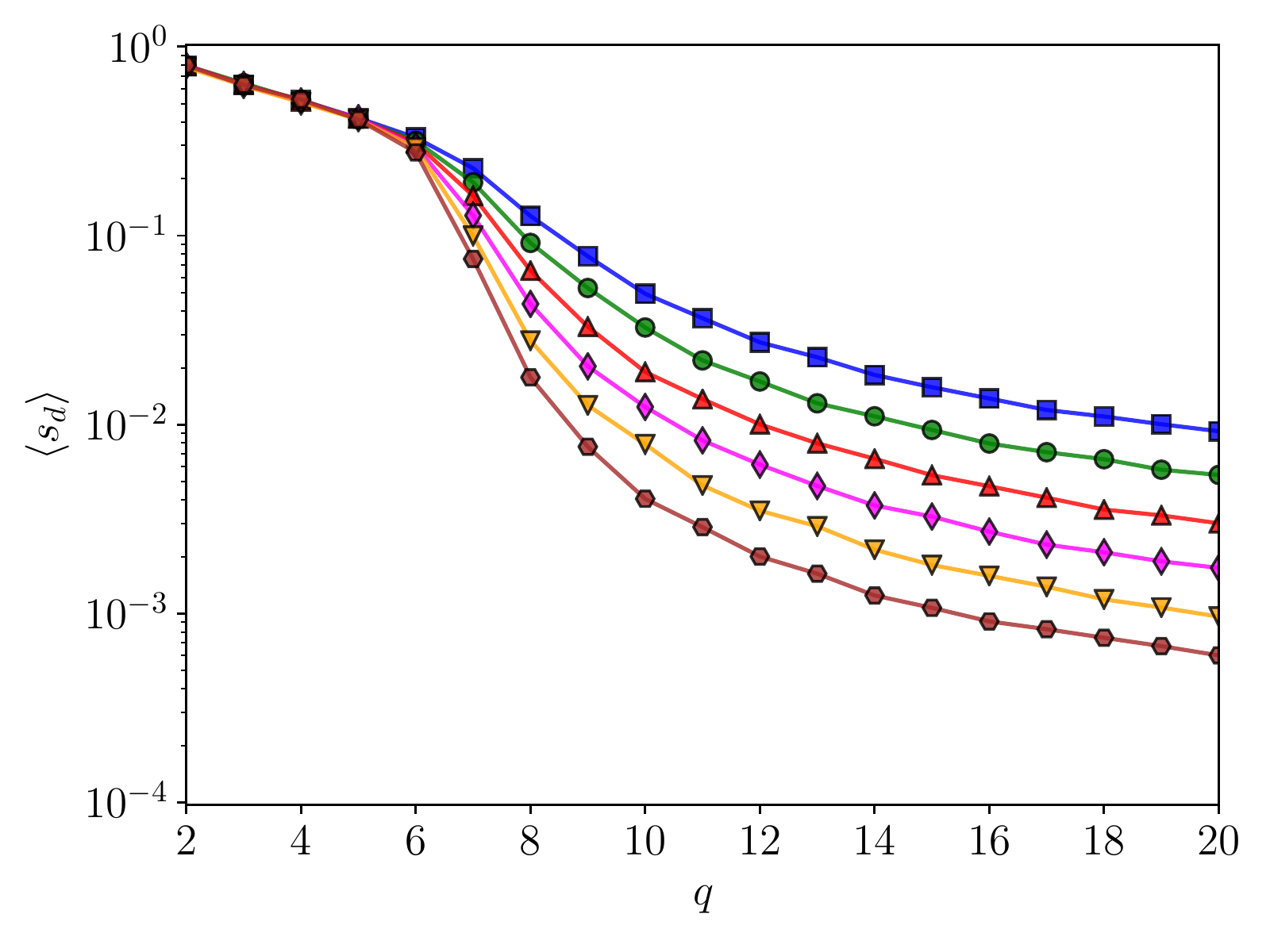} 
\includegraphics[width=0.48\textwidth]{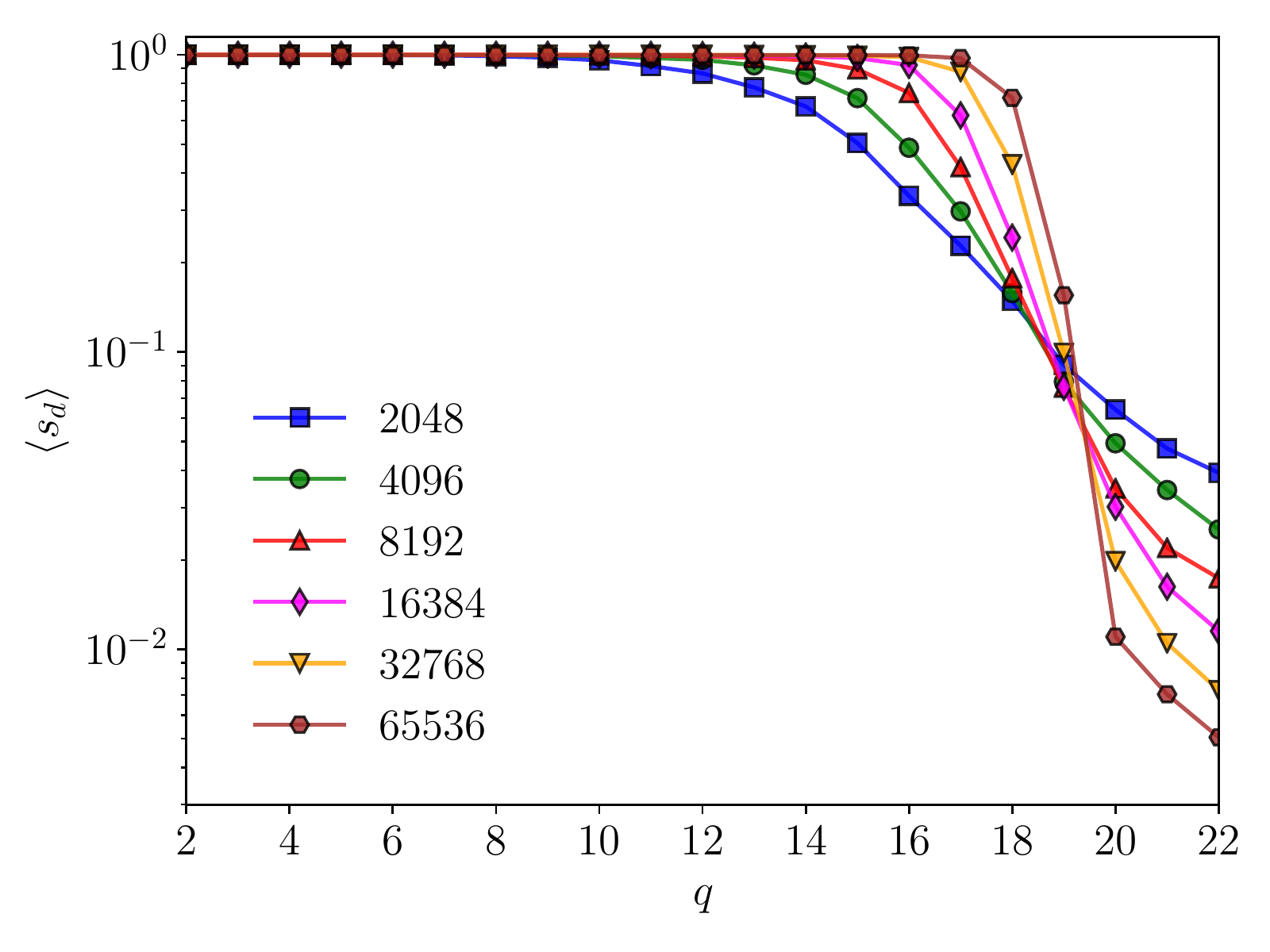} 
\caption{Mean fraction of agents in the largest cultural domain  $\langle s_d \rangle $   as function of the initial  cultural diversity $q$ for $F=2$ (upper panel) and $F=3$ (lower panel).  The  agents are fixed at the sites of  random geometric graphs (RGGs), i.e., the mobility parameter is $\delta =0$.  The continuous phase transition for $F=2$ takes place between $q=6$ and $q=7$, whereas the discontinuous phase transition for  $F=3$ takes place between $q=19$ and $q=20$. The system sizes are $N= 2^l$ with $ l=11, \ldots, 16$, as indicated.
 }  
\label{fig:2}  
\end{figure}

Because the Axelrod model was not studied in the case the agents are fixed at the sites of   RGGs, it is instructive to consider briefly the static limit, $\delta =0$. Figure \ref{fig:2} shows  the dependence of  the fraction of agents in the largest cultural domain $\left \langle s_d \right \rangle  $ on the initial diversity of the system $q$ for $F=2$ (upper panel) and $F=3$ (lower panel). This figure  exhibits the hallmark of Axelrod's model for static agents, namely, the existence of a phase transition  between 
ordered  absorbing configurations, which  are characterized by the presence of few cultural domains of macroscopic size (i.e., $\left \langle s_d \right \rangle$  is nonzero for $N \to \infty$),   and disordered absorbing configurations, where all cultural domains are microscopic (i.e., $\left \langle s_d \right \rangle \to 0$  for $N \to \infty$).   The phase transition  is continuous for $F=2$ and discontinuous for $F =3$, similarly to the results for the square lattice \cite{Castellano_00,Peres_15}. 
We note that the existence of the ordered phase  implies a symmetry breaking leading to the dominance of few cultures, which does not  happen for the one-dimensional model that, for $F=2$, exhibits  disordered  absorbing configurations for all $q$  \cite{Vilone_02,Biral_15}.

For the RGG, our results indicate  that the continuous transition  ($F=2$) takes place between $q=6$ and $q=7$ and the  discontinuous transition  ($F=3$) takes place between $q=19$ and $q=20$. We note that there is no need to use sophisticated methods (e.g., Binder cumulants \cite{Binder_81}) to determine the critical point $q_c$, since we can only state that  $q^*  \leq q_c  \leq q^* +1$, where $q^*$ can be determined with almost certainty using the results of Fig.\ \ref{fig:2}. For instance, $q^*= 6$ for $F=2$ and
$q^*= 19$ for $F=3$. In the static scenario, $q^*$ is the largest $q$ for which $\left \langle s_d \right \rangle > 0$ in the thermodynamic limit.   The order of the phase transition is determined by the presence or not of  crossings of  the curves of the order parameter $\left \langle s_d \right \rangle  $ vs.\  $q$ for different system sizes.

Since the RGG is almost surely connected for $\alpha =1.75$,  we have $\left \langle s_c \right \rangle  \to 1$  and $\left \langle n_c \right \rangle \to  0 $ for $N \to \infty$, regardless of the value of $q$. We recall that  the topology of the influence network is not influenced by the social dynamics for $\delta=0$. 

%-----------------------------------------------------
%-----------------------------------------------------

%-----------------------------------------------------
\begin{figure} 
\centering  
\includegraphics[width=0.48\textwidth]{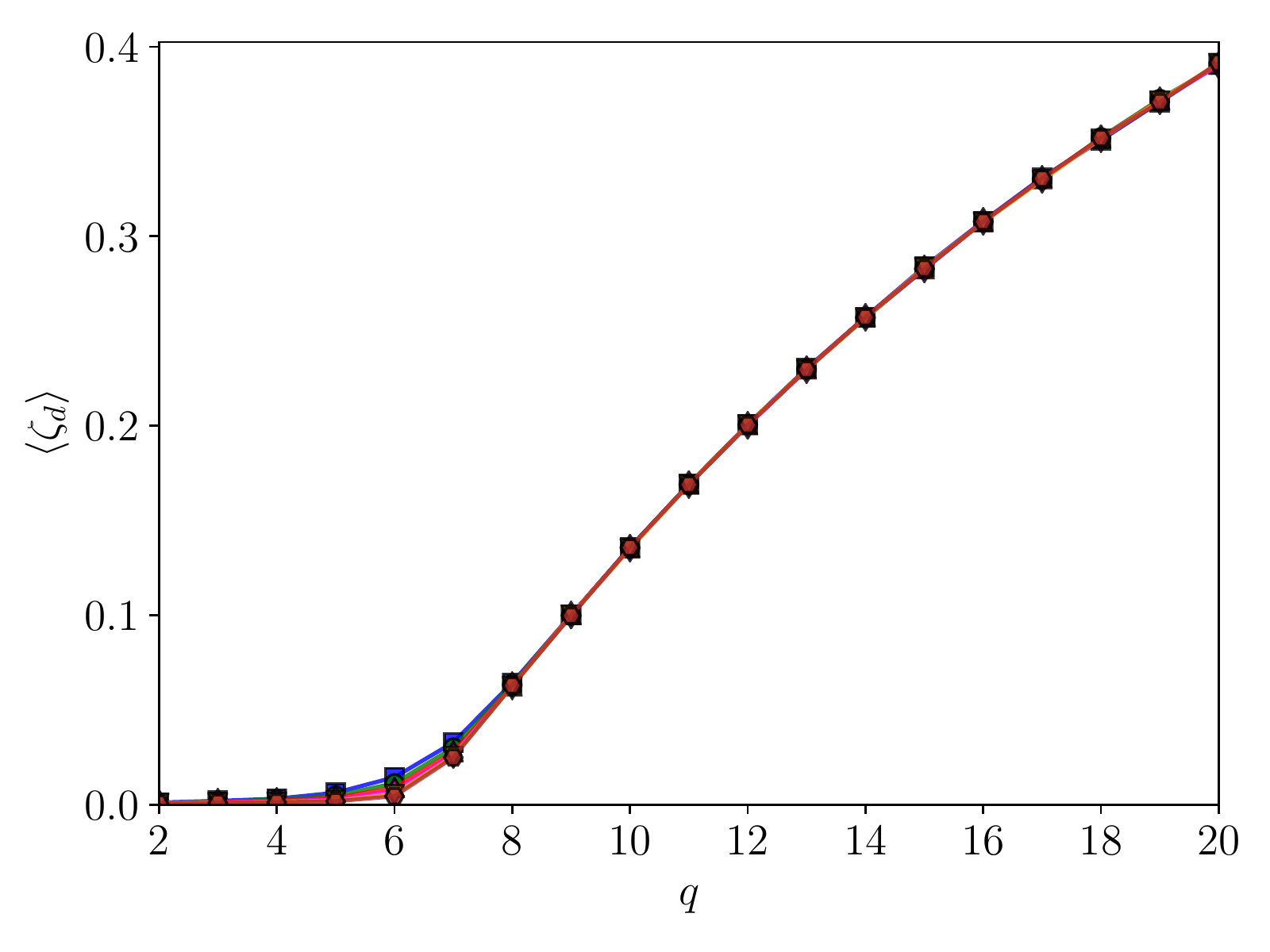} 
\includegraphics[width=0.48\textwidth]{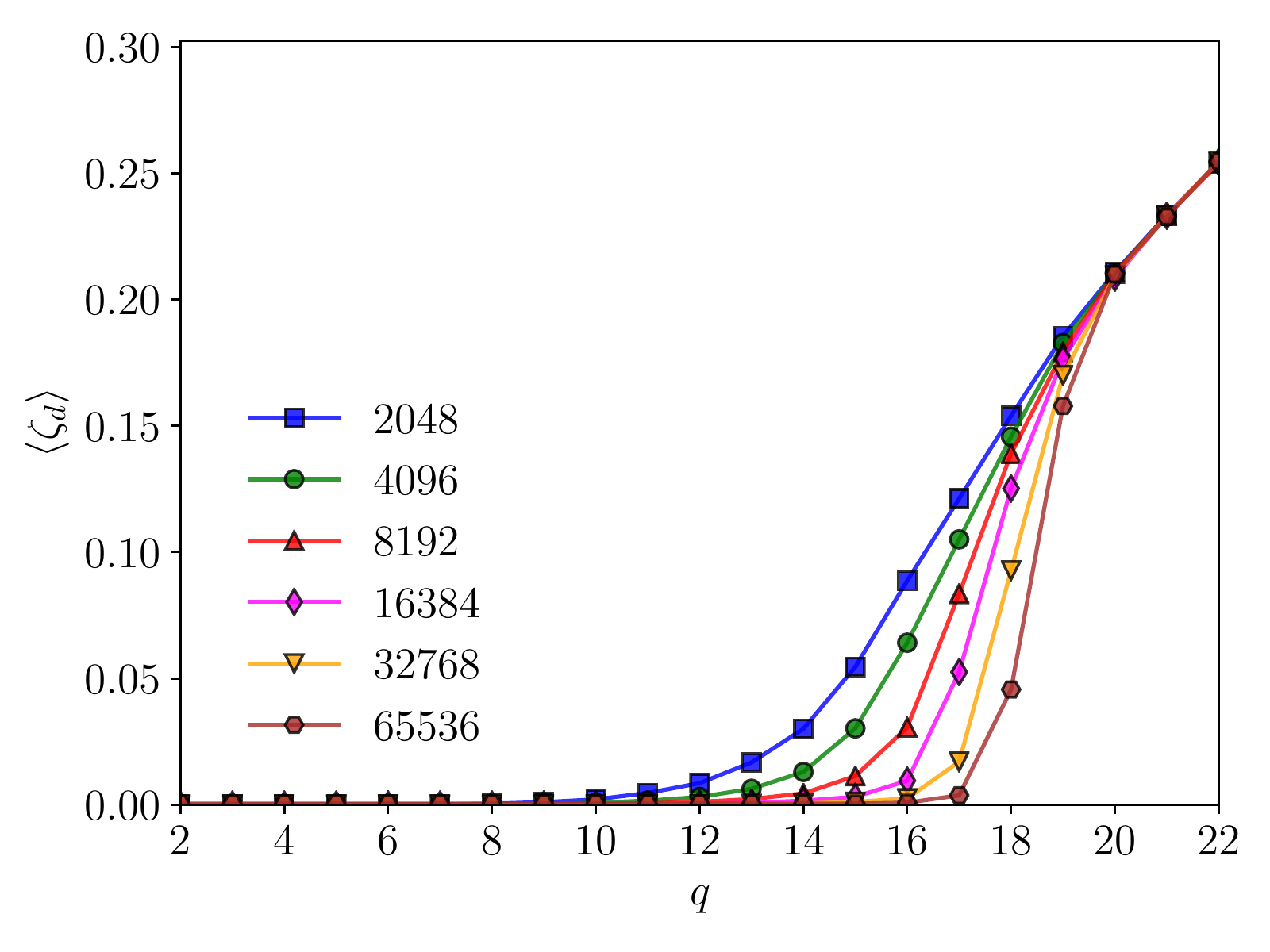} 
\caption{Mean fraction of culturally isolated agents $\langle \zeta_d \rangle $  as function of the initial  cultural diversity $q$ for $F=2$ (upper panel) and $F=3$ (lower panel).  The  agents are fixed at the sites of  random geometric graphs (RGGs). i.e.,  $\delta =0$.  The system sizes are $N= 2^l$ with $ l=11, \ldots, 16$ as indicated.
 }  
\label{fig:3}  
\end{figure}
%-----------------------------------------------------

Figure  \ref{fig:3} shows 
the mean fraction of culturally isolated agents $\langle \zeta_d \rangle $ for $F=2$ and $F=3$.  Somewhat surprisingly, a  large proportion of the agents are culturally isolated (i.e.,  they do not  share any cultural feature with the agents in their influence neighborhoods) in the disordered regime of the static limit $\delta=0$. According to our definition, those agents are uncomfortable and would move with certainty if they were allowed to.  We note that  $\langle \zeta_d \rangle \to 0$ for $N \to \infty$ in the ordered regime.  We advance that for $N$ and $q$ such that $NF/q \gg 1$, the comfort-driven mobility  is very effective to suppress  culturally isolated agents in the absorbing configurations (see \ref{App_q}).

\subsection{Mobile  agents for fixed system size}
\begin{figure}
\centering  
 \subfigure{\includegraphics[width=0.48\textwidth]{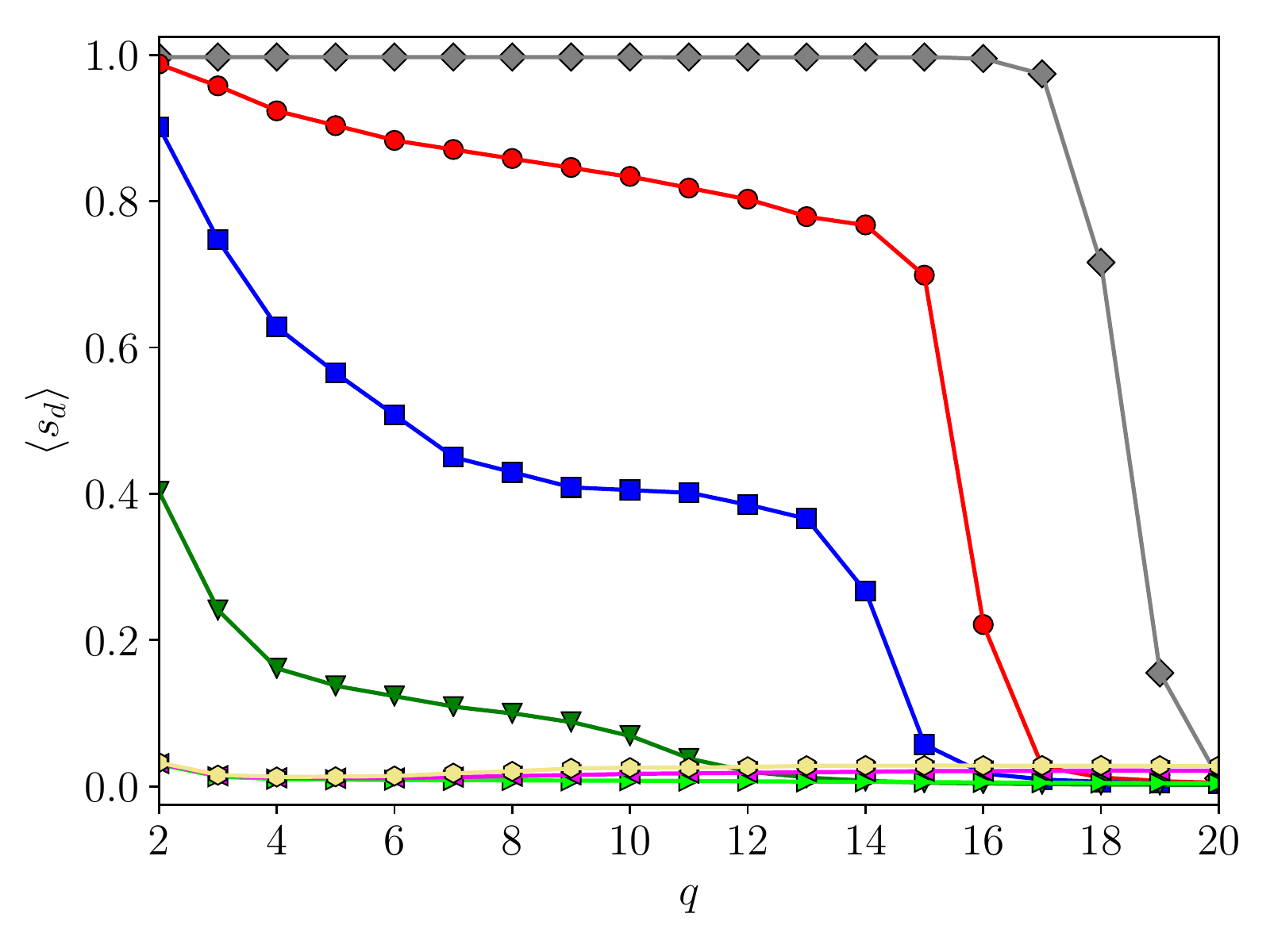}} 
 \subfigure{\includegraphics[width=0.48\textwidth]{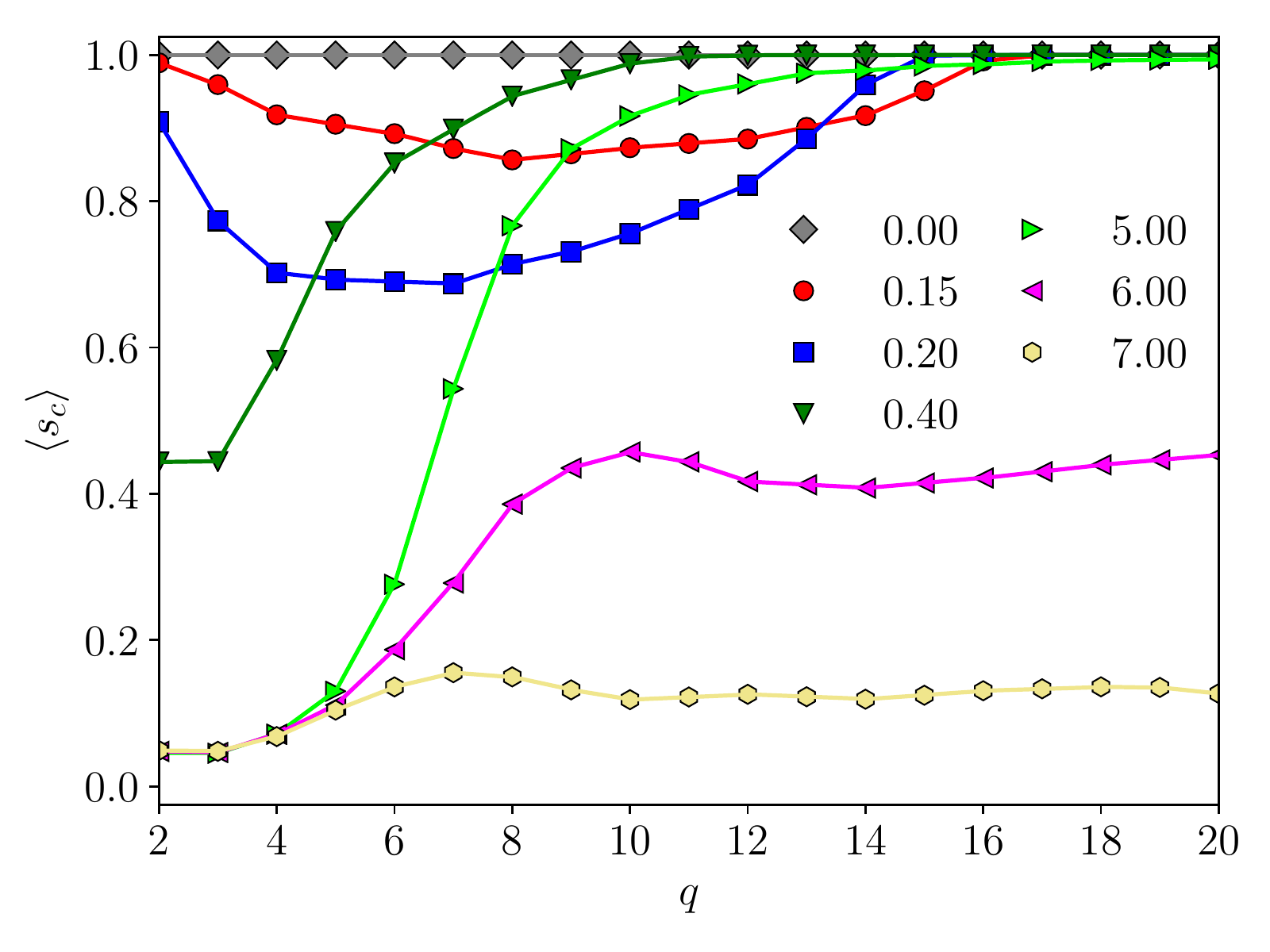}}
\caption{Mean fraction of agents in the largest cultural domain  $\langle s_d \rangle $ (upper panel)  and in the largest component of the influence network  $\langle s_c \rangle $ (lower panel)  as functions  of the initial  cultural diversity $q$ for $N= 2^{16}$ and step sizes $\delta =0, 0.15, 0.2, 0.4, 5, 6$ and  $7$, as indicated.
 }  
\label{fig:4}  
\end{figure}

Here we investigate the effects of the mobility parameter $\delta$ on the properties of the absorbing configurations for the system size $N=2^{16}$. Accordingly, Fig.\ \ref{fig:4} summarizes the influence of    the initial diversity $q$ on   $\langle s_d \rangle $ and 
$\langle s_c \rangle $ for  several representative values of the  step size $\delta$. The  upper panel of the figure shows that  the size of the largest cultural domain  decreases with increasing $\delta$ and that the disruptive effect of the mobility parameter on $\langle s_d \rangle $    is enhanced for large $q$. These findings  are consistent with the  expectation that the comfort-driven mobility should increase cultural diversity (i.e.,  decrease $\langle s_d \rangle $) since it reduces the strength of social influence  by decreasing the odds of repeated interactions between  the same pair of agents. We advance, however,  that our analysis of the finite size effects will show that $\langle s_d \rangle  \to 0$ for all $q$ and $\delta > 0$ in the thermodynamic limit, so the statistics of cultural domains is not  informative in the context of mobile agents. 

%-----------------------------------------------------
%---------------

The low informative power  of $\langle s_d \rangle$ contrasts with the mean size of the largest component $\langle s_c \rangle $, which is shown in  the lower panel of Fig.\ \ref{fig:4} and highlights the  strong influence of the model parameters  on  the  connectedness of the influence network, manifested by the non-monotonous dependence  of $\langle s_c \rangle $ on both $q$ and $\delta>0$.  Recalling
that $\langle s_c \rangle $ offers a picture of the fragmentation of the  influence network, this panel shows that  for small step sizes $\delta$   the fragmentation is more severe  for low  and intermediate values of the initial cultural diversity  and that the network is almost connected for  large $q$.  For $\delta > 6$, the influence network  is severely fragmented regardless of the value of $q$. Interestingly, although   the fragmentation
of the influence network results  necessarily in a decrease of the size of the largest cultural domain since $\langle s_c \rangle \geq \langle s_d \rangle $, the size of the largest cultural  domain  is  actually smaller  when the network is almost connected 
(i.e., $\langle s_c \rangle \approx 1$) than when it is severely  fragmented.

\subsection{Finite size effects} \label{FSE}

Here we argue that the comfort-driven mobility  induces a   fragmentation  transition separating the regime where all components are microscopic (i.e., $\langle s_c \rangle \to 0 $ for $N \to \infty$) from the regime where at least one component is macroscopic (i.e., $\langle s_c \rangle > 0 $ for $N \to \infty$). Henceforth we will refer to these regimes as the  severely and mildly fragmented regimes.

\begin{figure}
\centering  
 \subfigure{\includegraphics[width=0.48\textwidth]{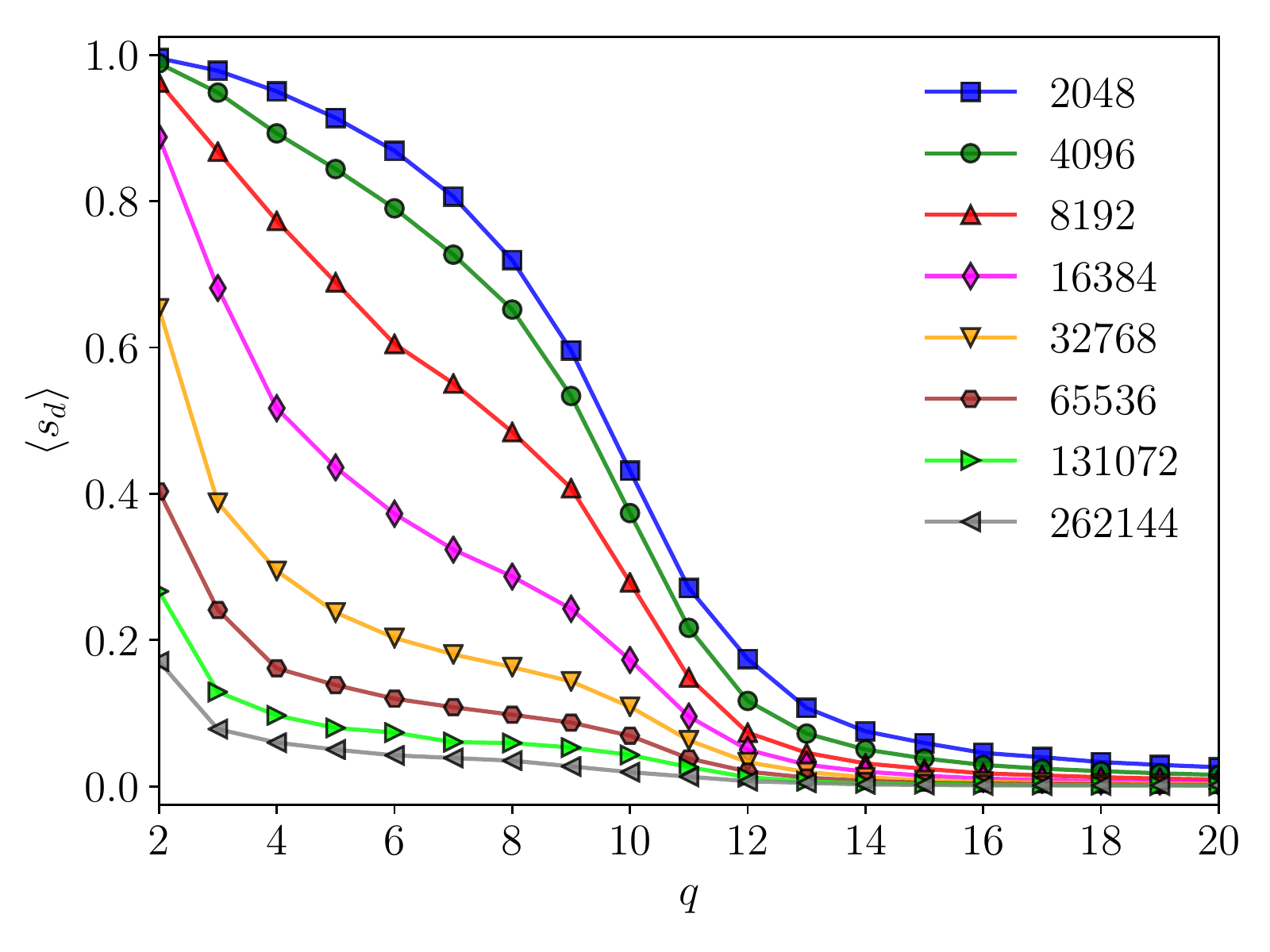}} 
\subfigure{\includegraphics[width=0.48\textwidth]{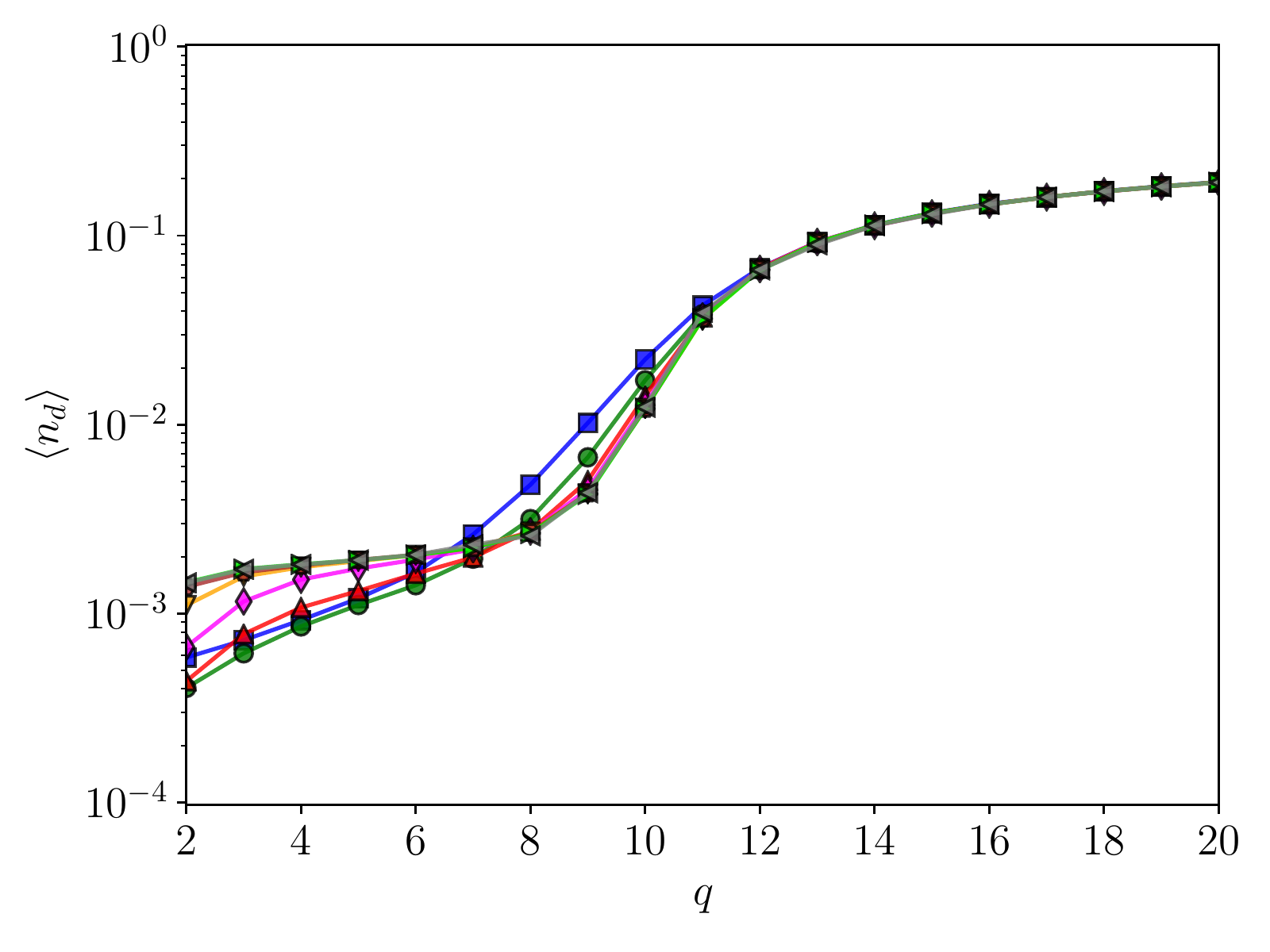}}
\caption{Mean fraction of agents in the largest cultural domain  $\langle s_d \rangle $ (upper panel),  and mean density of  cultural domains $\langle n_d \rangle $ (lower panel)
 as functions  of the initial  cultural diversity $q$ for $\delta =0.4$ and  $N= 2^l$ with $ l=11, \ldots, 18$,  as indicated. 
 }  
\label{fig:5}  
\end{figure}

\begin{figure}
\centering  
 \subfigure{\includegraphics[width=0.48\textwidth]{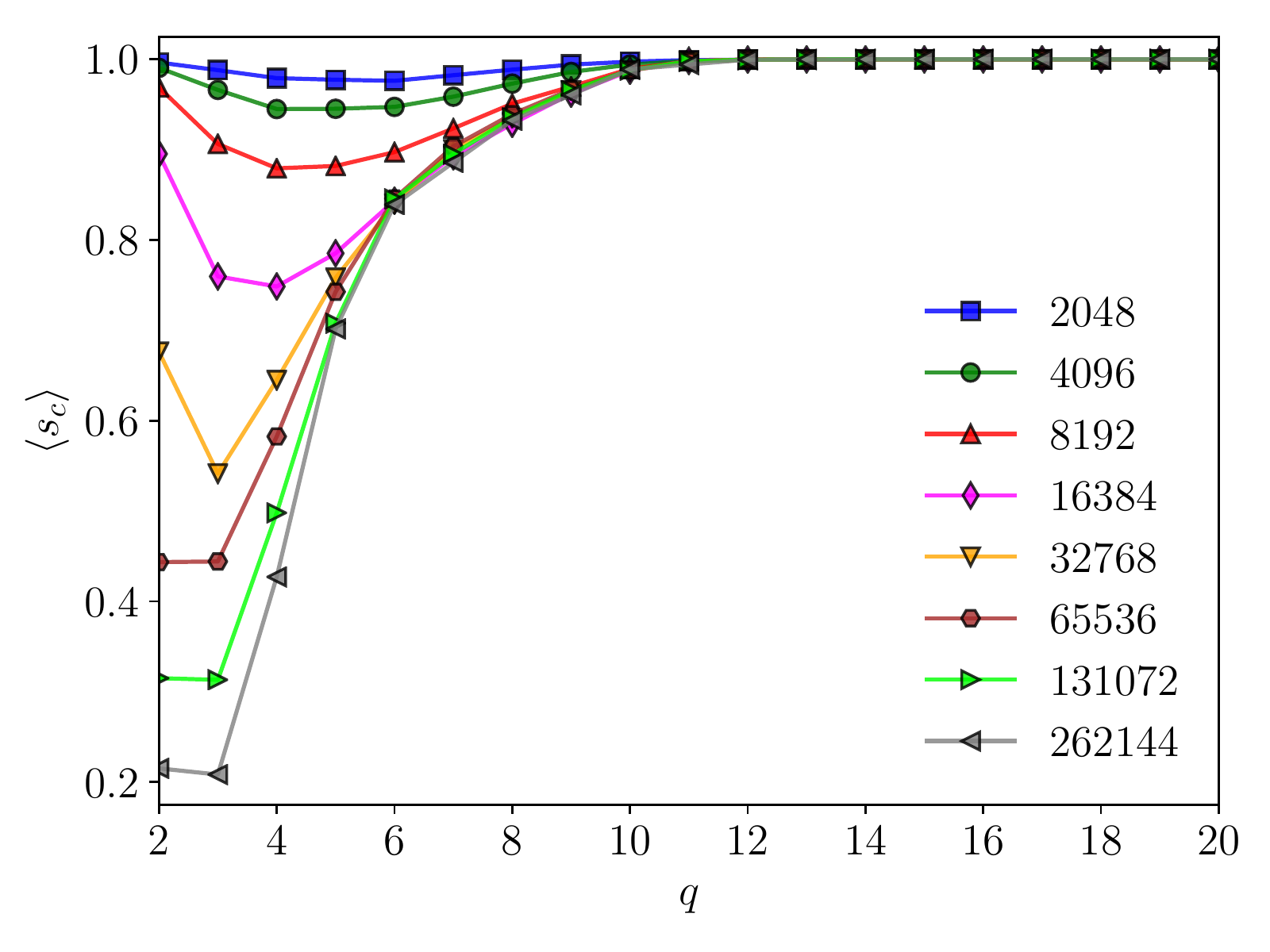}} 
\subfigure{\includegraphics[width=0.48\textwidth]{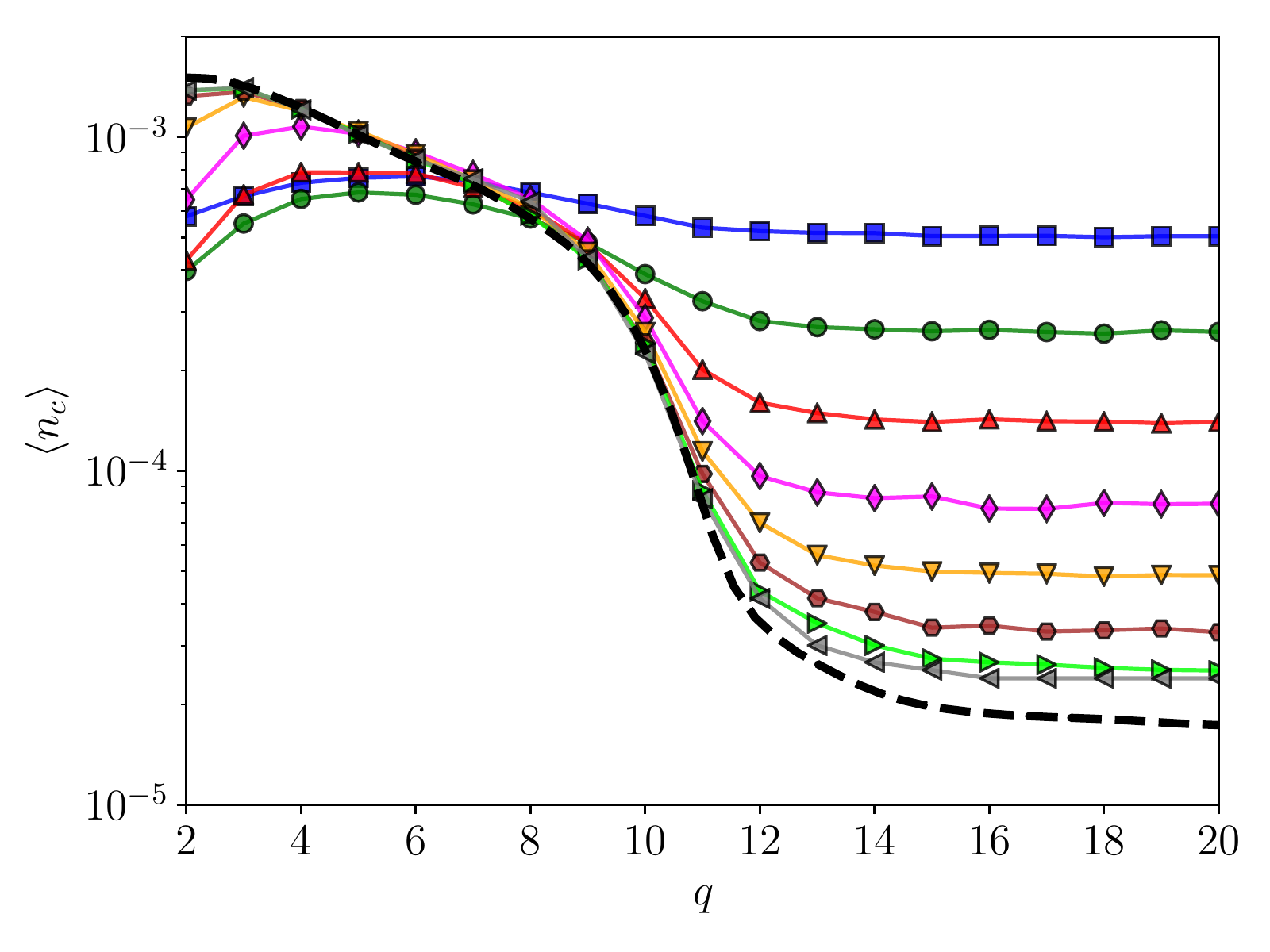}}
\caption{Mean  fraction of agents in the largest component of the influence network  $\langle s_c \rangle $ (upper panel) and mean density of components $\langle n_c \rangle $ (lower panel)
 as functions  of the initial  cultural diversity $q$ for $\delta =0.4$ and  $N= 2^l$ with $ l=11, \ldots, 18$,  as indicated. The dashed curve in the
lower panel is  the fitting parameter $\langle n_c \rangle_\infty$ of  the extrapolation of $\langle n_c \rangle $ to  $N \to \infty$. The  continuous transition separating the regime where the largest component of the influence network is macroscopic from the regime where all components are microscopic 
 takes place between $q = 5$ and  $q = 6$.
 }  
\label{fig:6}  
\end{figure}

Figures \ref{fig:5} and \ref{fig:6}  summarize our results for $\delta=0.4$. In particular, the upper panel of Fig. \ref{fig:5} shows that $\langle s_d \rangle \to 0$ as $N$ increases, regardless of the value of $q$. More pointedly,  for large $N$ we find that  $\langle s_d \rangle$ vanishes as $N^{-0.8}$ for all $q$ (data not shown). The lower panel of this figure shows that the density of cultural domains $\langle n_d \rangle$ tends rapidly to its smooth asymptotic limit $\langle n_d \rangle_\infty > 0$, indicating thus the presence of a macroscopic number of cultural domains   for all $q$.
Hence the comfort-driven mobility eliminates altogether the  ordered absorbing configurations of the static limit.   We recall that  those configurations are  the sole attractors of the dynamics  in the range  $q < 20$ for $\delta=0$ (see lower panel of Fig.\ \ref{fig:2}).
%Thus, the statistics of cultural domains are not useful to describe the system of mobile agents for large $N$. 

 %-----------------------------------------------------
%----------------------------------------------------- 

\begin{figure}
\centering  
\includegraphics[width=0.48\textwidth]{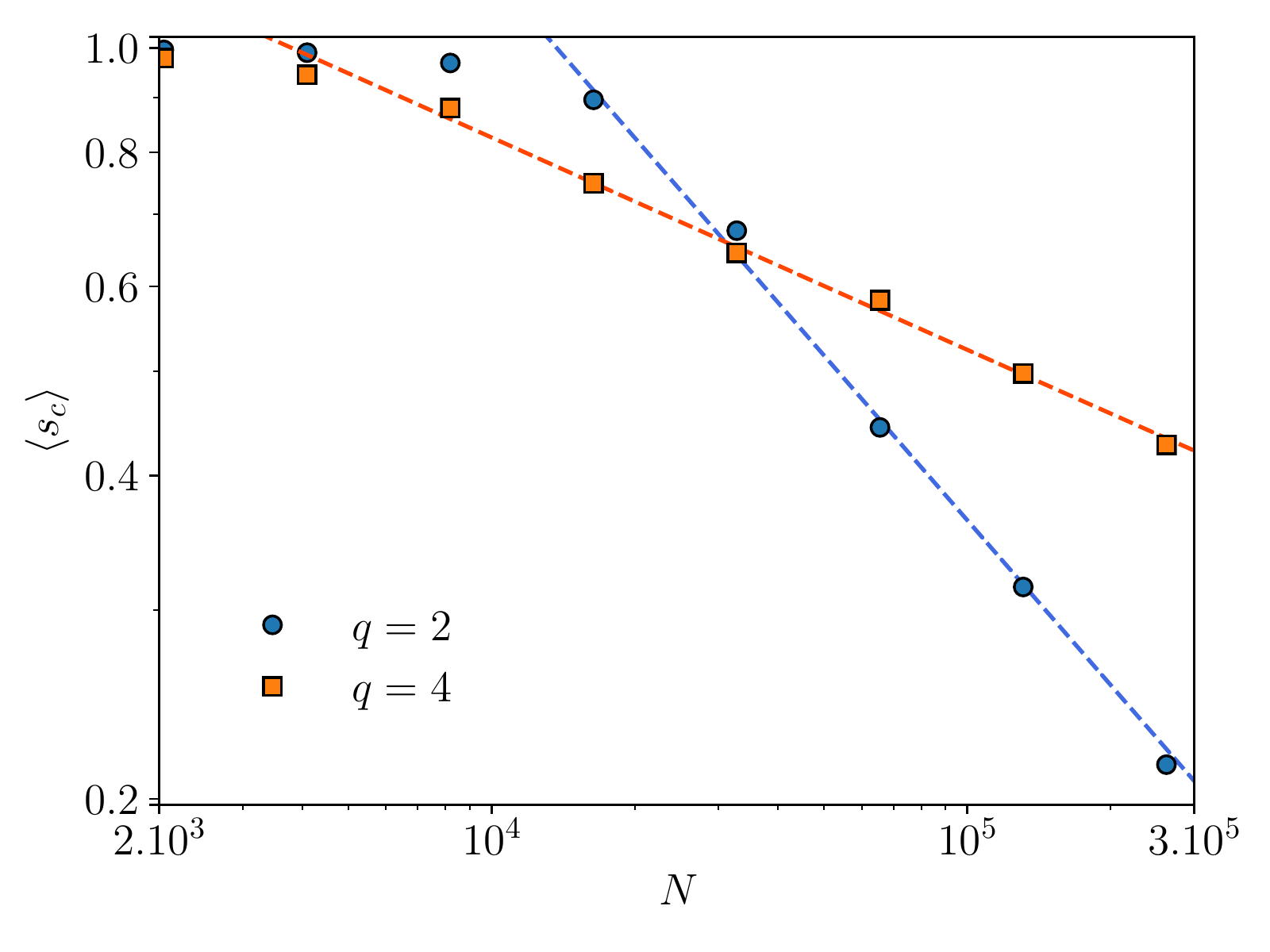} 
\caption{Mean fraction of agents in the largest component  $\langle s_c \rangle $ as function of the system size $N$ for  the step size $\delta=0.4$ and $q=2,4$, as indicated. The  dashed lines are the fittings  $\langle s_c \rangle = 135 N ^{-0.51} $ for $q=2$ and
$\langle s_c \rangle = 5.24 N ^{-0.20} $ for $q=4$.
 }  
\label{fig:7}  
\end{figure}

The  left  panel of Fig.\ \ref{fig:6} indicates that  in the thermodynamic limit  $\langle s_c \rangle$ exhibits a transition between  $q=5$ and $q=6$ that separates the regime where the influence network is severely fragmented (i.e., all components are microscopic) from the regime where that network exhibits a macroscopic component. For instance, for $q=4$ we find that $\langle s_c \rangle$ vanishes like the power law $N^{-0.2}$ with increasing $N$, whereas it
vanishes as $N^{-0.5}$ for $q=2$ (see Fig.\ \ref{fig:7}).

We note that since $\langle s_c \rangle < 1$  the influence network is always  fragmented for $\delta > 0$. The main  difference between the severely and the mildly fragmented  regimes is the presence or not of a macroscopic component in the thermodynamic limit. In addition, since  the ratio $\langle s_d \rangle/ \langle s_c \rangle $ tends to zero for large $N$ there is coexistence between  different
cultures  inside  the largest  component in both fragmentation regimes.

%-----------------------------------------------------
%----------------------------------------------------- 

%-----------------------------------------------------
%-----------------------------------------------------
   
The  right  panel of Fig.\ \ref{fig:6} shows that the mean density of  components $\langle n_c \rangle $ is not affected by the fragmentation transition revealed in the study of the largest component. In addition, it shows that   $\langle n_c \rangle > 0 $ in the limit $N \to \infty$, implying thus that  the macroscopic component  coexists with a macroscopic number of microscopic components
for $q \geq 6$.  This finding supports the claim that $\langle s_c \rangle < 1$  for $\delta > 0$.   From  our results it is not possible to tell whether there are other macroscopic components in the mildly fragmented regime besides the largest one. Of course, since in the severely fragmented regime  the largest component is microscopic (i.e.,  $\langle s_c \rangle \to 0$ for $N \to \infty$ ), so are all the other components.
The extrapolation of  $\langle n_c \rangle > 0 $   to
  $N \to \infty$ (dashed curve in the  right  panel of Fig.\ \ref{fig:6}) was obtained through the fitting  $\langle n_c \rangle  = \langle n_c \rangle_\infty + a_n/N$ where $\langle n_c \rangle_\infty$ and $a_n$ are fitting parameters that depend on $q$ and $\delta$. Most interestingly, the parameter $a_n$ changes sign at  about $q=6$, where the  fragmentation transition takes place.  Viewing $1/\langle n_c \rangle$  as an estimator of the average component size, we can infer that when $a_n$ is positive ($q \geq 6$), the components become  bigger as the system size increases, which is consistent with the presence of a macroscopic component in the mildly fragmented regime. In turn, when $a_n$ is negative  the components become  smaller as $N$ increases, which is consistent with the existence of  solely microscopic components in  the severely fragmented regime.

    The results for $\delta < 0.4$ are similar to those exhibited in
  Figs.\ \ref{fig:5} and \ref{fig:6} but,   as hinted at  in Fig.\ref{fig:4}, the finite size effects are very strong, requiring  the use of impracticably large system sizes to characterize the severely fragmented regime.
  
  %-----------------------------------------------------
\begin{figure} 
\centering  
 \subfigure{\includegraphics[width=0.48\textwidth]{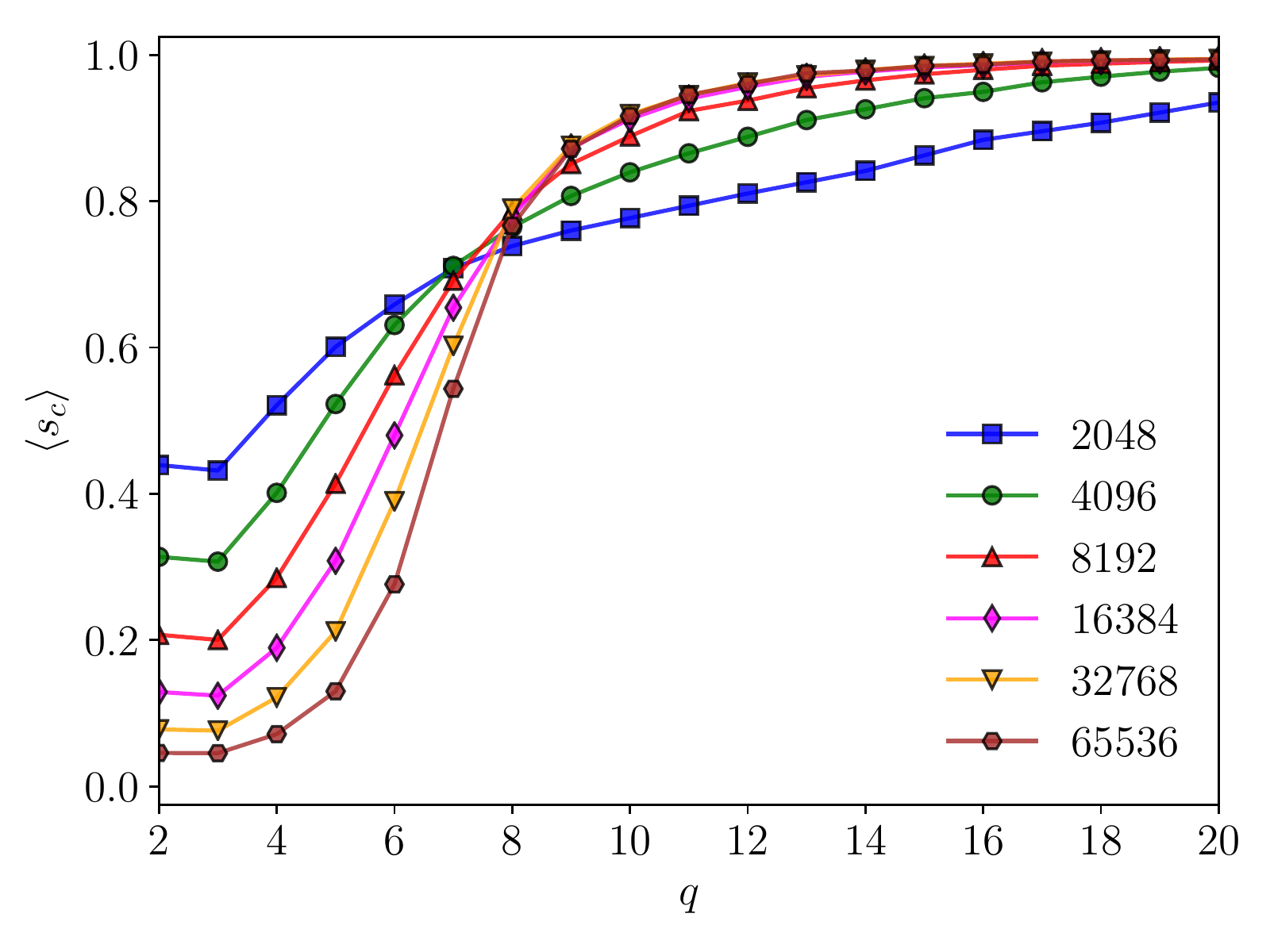}} 
 \subfigure{\includegraphics[width=0.48\textwidth]{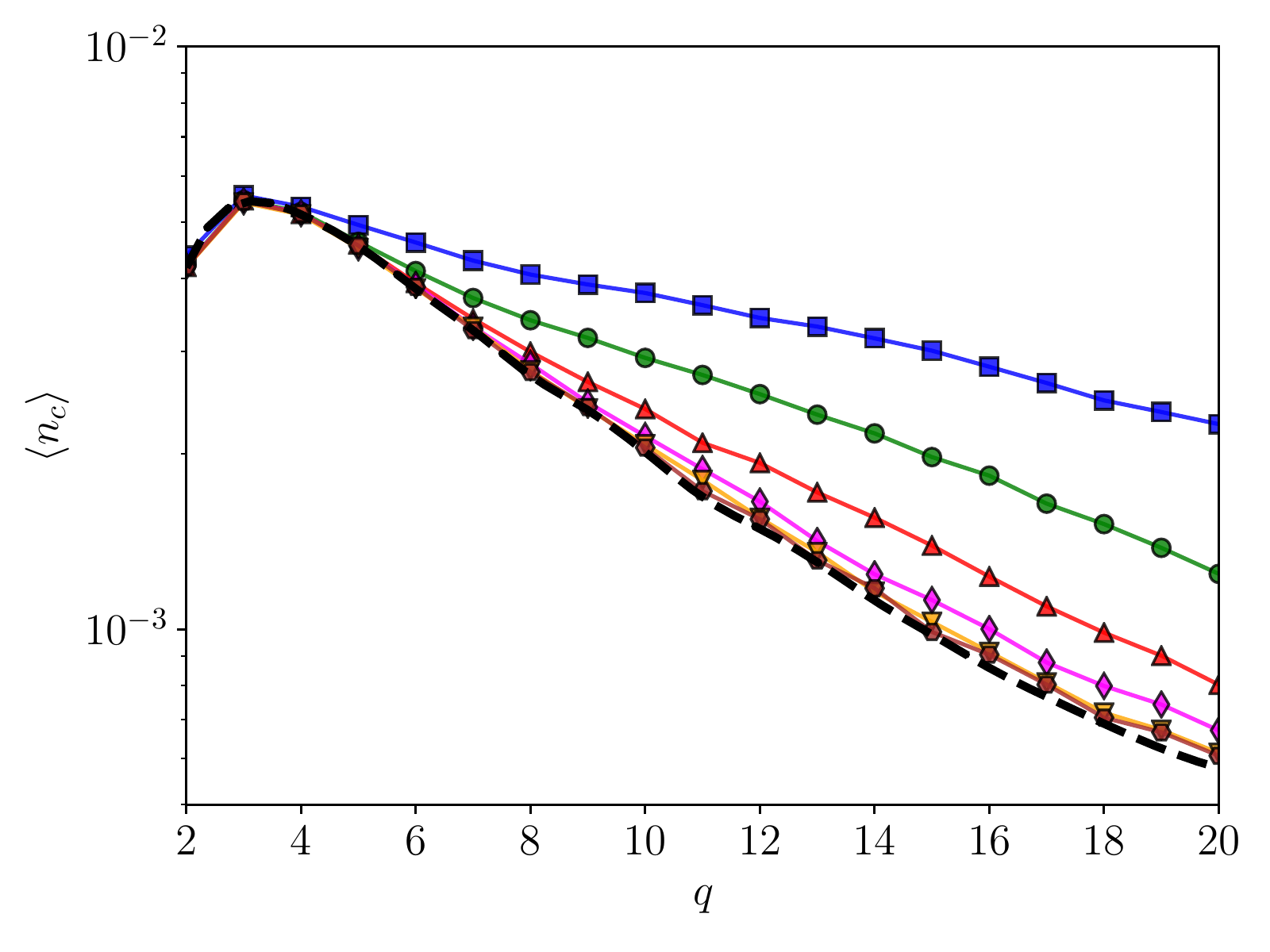}} 
 %\subfigure{\includegraphics[width=0.48\textwidth]{fig5c.pdf}} 
\caption{Mean fraction of agents in the largest component  $\langle s_c \rangle $  (upper panel) and mean density of components $\langle n_c \rangle $  (lower panel)  as functions  of the initial  cultural diversity $q$ for $\delta =5$
 and systems of size $N= 2^l$ with $ l=11, \ldots, 16$,  as indicated. The dashed curve in the
lower panel is  the fitting parameter $\langle n_c \rangle_\infty$ of the extrapolation of $\langle n_c \rangle $ to  $N \to \infty$. The  discontinuous  transition separating the regime where the largest component of the influence network is macroscopic from the regime where all components are microscopic 
 takes place between $q=7$ and $q = 8$.
 }  
\label{fig:8}  
\end{figure}
%-----------------------------------------------------

 The normalized mean size of the largest component of the influence network  $\langle s_c \rangle $ is the order parameter of the  fragmentation transition and its dependence on the system size $N$ may shed some light on the nature of the phase transition. In particular, it would be of interest to know whether the transition between the severely and mildly fragmented regimes is continuous or discontinuous.  Although it is somewhat problematic to discuss this classification in the case the independent variable $q$ is discrete, the fact that the curves of   $\langle s_c \rangle $ vs.\ $q$ for different $N$ do not cross (left  panel of Fig.\ \ref{fig:6}) suggests that the fragmentation transition is continuous for $\delta=0.4$.  It is interesting that the  curves of   $\langle n_c \rangle $ for distinct system sizes do cross at about $q=6$ (right  panel of Fig.\ \ref{fig:6}) but since  $\langle n_c \rangle $ is not an order parameter (it is nonzero in both fragmentation regimes),  it offers no information on the nature of the transition. 
 These findings contrast starkly with the results for $\delta=5$ shown in   Fig.\ \ref{fig:8}, where the crossing of the curves for different system sizes happens for  $\langle s_c \rangle $ but not for $\langle n_c \rangle $. This is the typical scenario of a discontinuous transition where the  order parameter becomes  independent of the system size at the threshold parameter $q_c$, which is thus determined by the intersection of the curves of $\langle s_c \rangle $ for large $N$. Since the relevant asymptotic behaviors, namely,   $\langle n_c \rangle > 0$ in the mildly fragmented regime and $\langle s_c \rangle \to 0$ in the severely fragmented regime, are more easily observed  in Fig.\ \ref{fig:8} than in Fig.\ \ref{fig:6}, we have not simulated the system sizes $N=2^{17}$ and $N=2^{18}$ for $\delta=5$. 
 
We note that  the nature of the fragmentation  transition is determined by the dependence of $\langle s_c \rangle $ on  $N$ 
in the mildly fragmented  regime. For instance, in that regime $\langle s_c \rangle $ decreases with increasing $N$ for $\delta=0.4$
(see upper panel of  Fig. \ref{fig:6}), whereas it increases with increasing $N$ for $\delta=5$ (see upper panel of  Fig. \ref{fig:8}). In fact, in the mildly fragmented  regime we can write
 $\langle s_c \rangle  = \langle s_c \rangle_\infty + a_s/N$, where $\langle s_c \rangle_\infty$ and $a_s$ are fitting parameters, so we can use the sign of the parameter $a_s$ to
 determine  whether the transition is continuous or discontinuous. Accordingly, we find that  $a_s$  changes sign at  $\delta \approx 4.2$, signaling thus a change on the nature of the fragmentation transition.

%-----------------------------------------------------
\begin{figure}[b!]
\centering  
\subfigure{\includegraphics[width=0.48\textwidth]{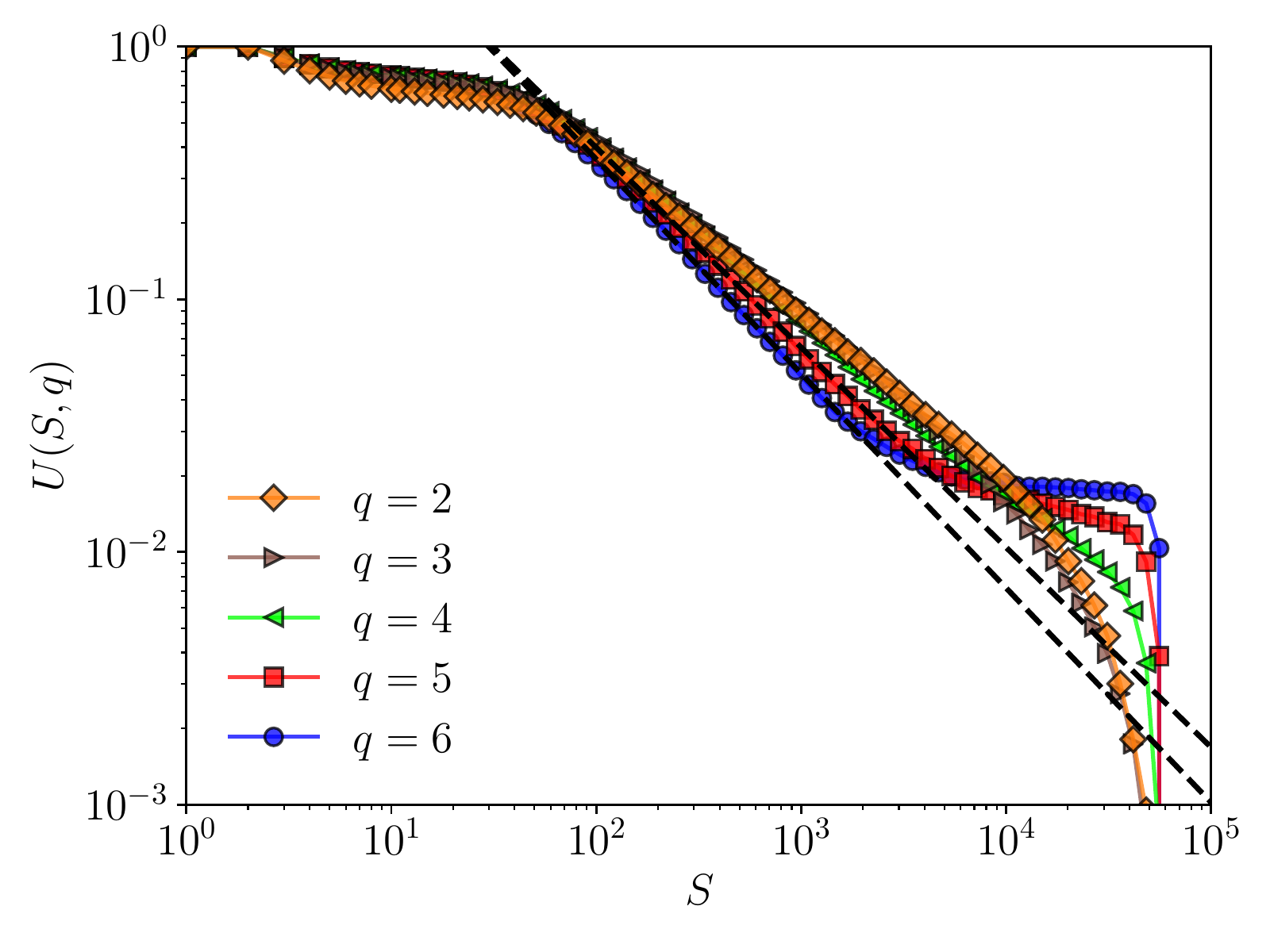}} 
 \subfigure{\includegraphics[width=0.48\textwidth]{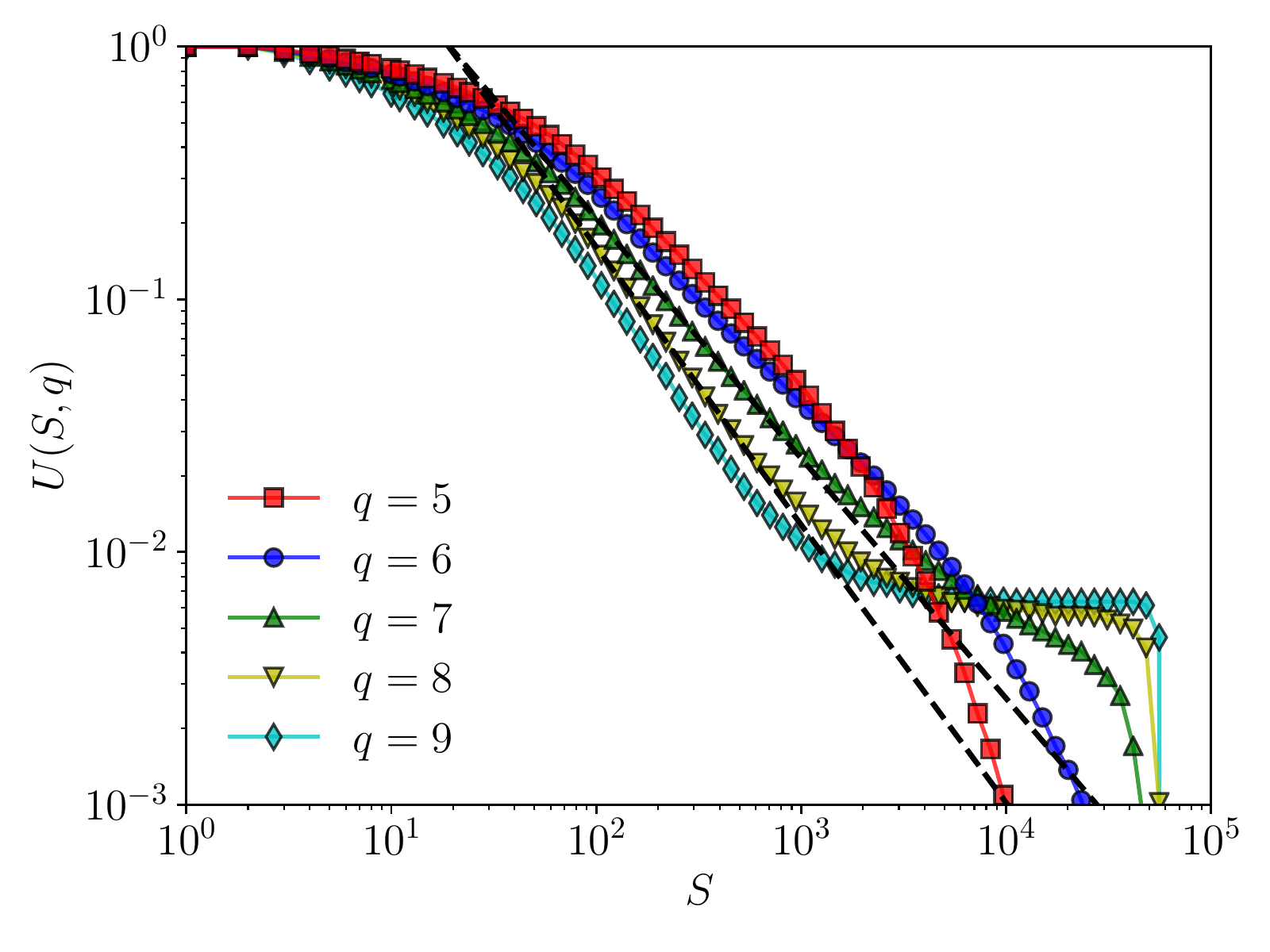}}
\caption{Fraction of components  of size larger than $S$ for $\delta = 0.4$ (upper panel)   and $\delta=5$  (lower panel). The  system size is $N=2^{16}$ and the initial cultural diversities $q$ are as indicated. The dashed  lines are the  power-law fittings for  $q=q^*$ and $q = q^* +1$ used to estimate the critical exponent $\tau$.
 }  
\label{fig:9}  
\end{figure}
%-----------------------------------------------------

 Now we consider an alternative method to estimate the transition points $q_c$, which  relies on the  analysis of the asymptotic behavior of the distribution of the component sizes \cite{Castellano_00}. In particular,  in Fig.\ \ref{fig:9}   we  show the cumulated distribution $U \left ( S,q \right )$, which gives   the fraction of components  of size larger than $S \in \left  [ 1,N \right ]$. In the limits of  large $N$ and $S$ such that $S \ll N$, we expect that $U \left ( S,q < q_c \right ) \to 0$ and $U \left ( S,q > q_c \right ) \to \mbox{cte} > 0$, so we can identify $q_c$ simply by observing the asymptotic behavior of this cumulated distribution. This approach yields the  same estimates as those based on the order parameter $\langle s_s \rangle $, viz., that for $\delta = 0.4$ the transition takes place between $q=5$ and $q=6$ (i.e., $q^* =5$) and that for $\delta = 5$ it occurs between $q=7$ and $q=8$ (i.e., $q^* =7$). More importantly, however, is the fact that $U \left ( S, q_c \right )$  decays as a power law $S^{1-\tau}$  and that the value of the exponent $\tau$ yields information on the nature of the transition at $q_c$ \cite{Castellano_00}.  More pointedly, the transition is continuous if $\tau \leq 2$ and discontinuous otherwise. The difficulty here is that because $q$ is discrete  we cannot determine $q_c$ in order to observe the power decay. Nevertheless, in the regions where $U \left ( S,q \right )$  at $q=q^*$ and $q=q^* + 1$ can be approximate by power laws we find  $\tau \approx 1.82$ for $\delta = 0.4$ and $\tau \approx 2.05$ for $\delta = 5$, which is consistent with the  conclusions based on  the crossing, not crossing of the curves showing system size effects. These estimates of the exponent $\tau$ are obtained by averaging the values of the slopes of the dashed straight lines  for $q=q^*$ and $q=q^* +1$ shown in Fig.\ \ref{fig:9}.

\begin{figure} 
\centering  
 \subfigure{\includegraphics[width=0.48\textwidth]{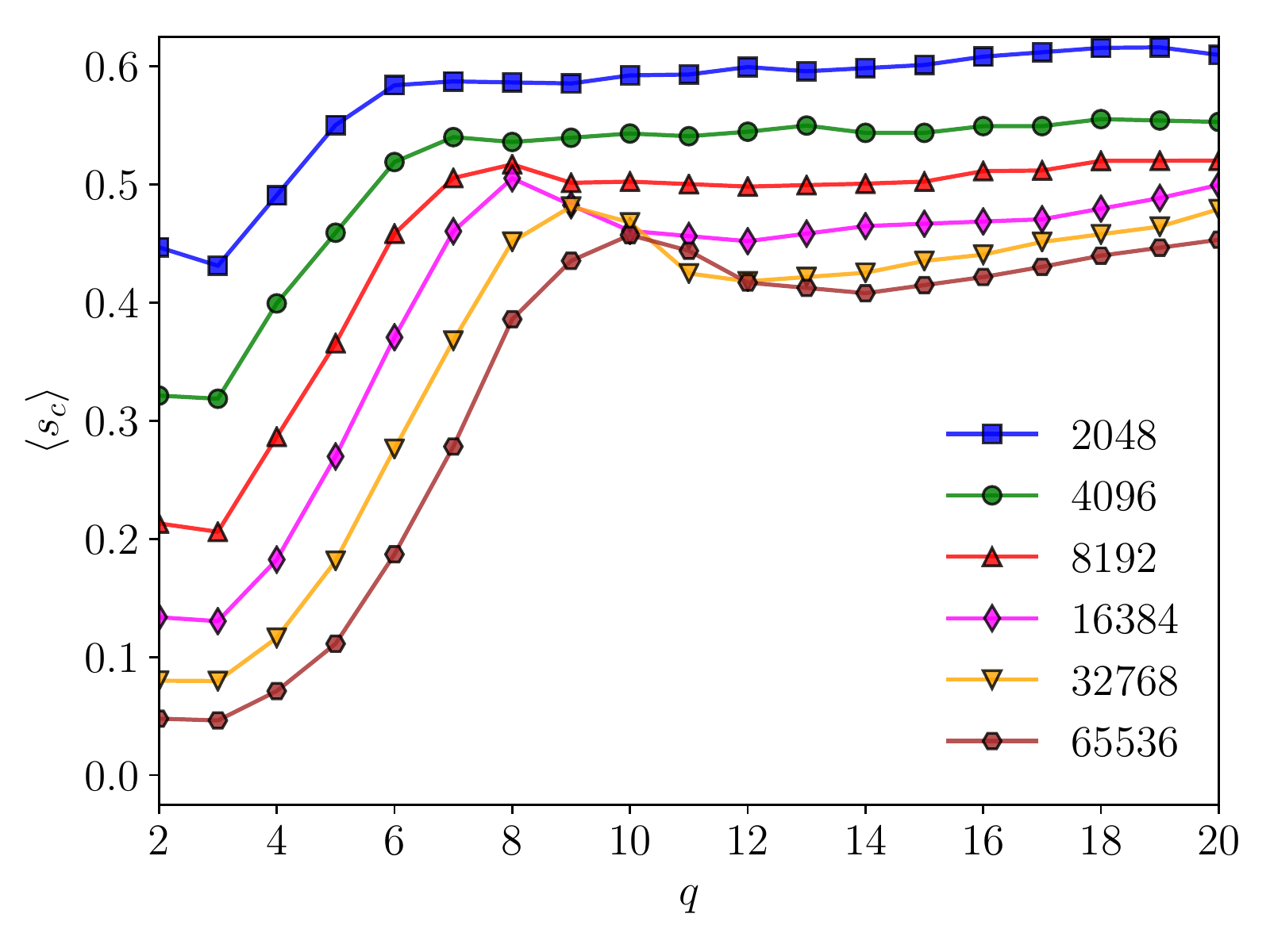}} 
 \subfigure{\includegraphics[width=0.48\textwidth]{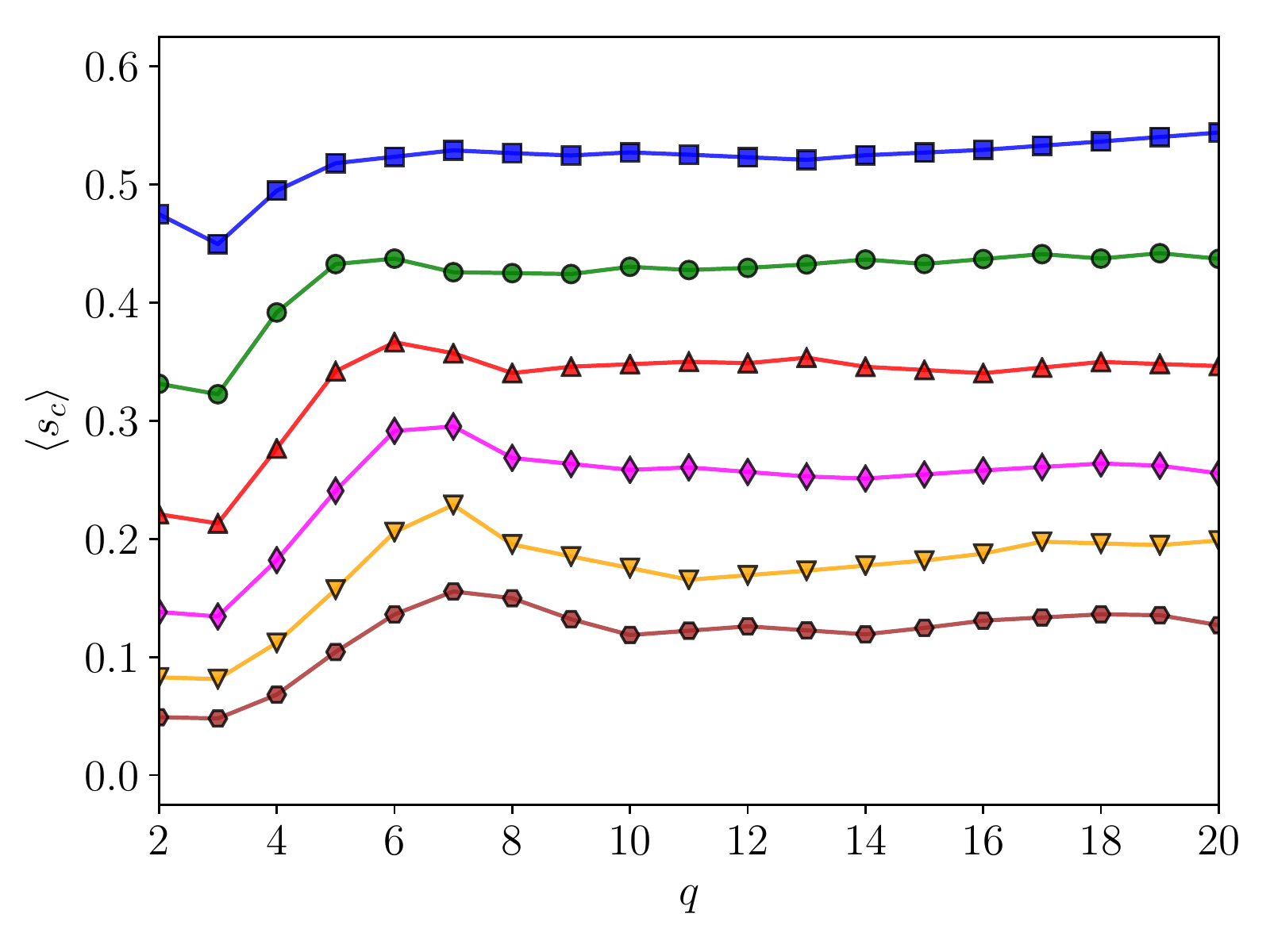}} 
 %\subfigure{\includegraphics[width=0.48\textwidth]{fig5c.pdf}} 
\caption{Mean fraction of agents in the largest component  $\langle s_c \rangle $ as function  of the initial  cultural diversity $q$ for $\delta =6$
 (upper panel),   $\delta =7$ (lower panel) and systems of size $N= 2^l$ with $ l=11, \ldots, 16$,  as indicated.
 }  
\label{fig:10}  
\end{figure}

\begin{figure}
\centering  
\includegraphics[width=0.48\textwidth]{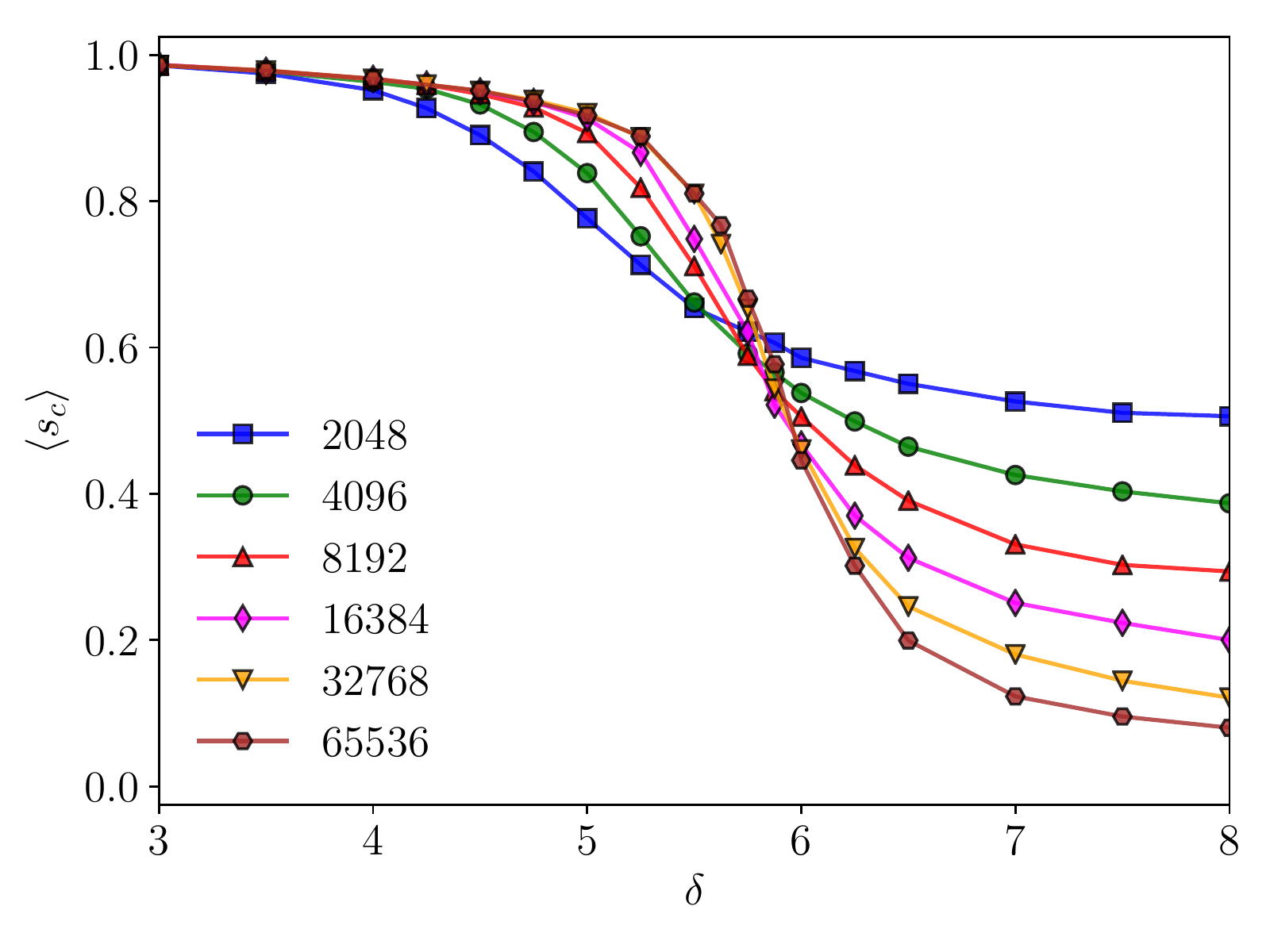} 
\caption{Mean fraction of agents in the largest component  $\langle s_c \rangle $ as function of the step size $\delta$  for $q=10$ and systems of size $N= 2^l$ with $ l=11, \ldots, 16$,  as indicated. The  discontinuous transition separating the regime where the largest component of the influence network is macroscopic from the regime where all components are microscopic 
 takes place at $\delta \approx 5.9$. 
 }  
\label{fig:11}  
\end{figure}

As hinted at in Fig.\ \ref{fig:4}, for   $\delta = 6$ and $7$,  the absorbing configurations remain highly fragmented even for large $q$. In fact, Fig.\ \ref{fig:10} summarizes the effects of the system size for these step sizes and shows that the absorbing configurations are  severely fragmented in the  thermodynamic limit, regardless of the value of $q$. This means that  the transition between the severely and mildly fragmented regimes disappears altogether at a  point $\delta^*$, so that for $\delta > \delta^*$ we have
$\langle s_c \rangle \to 0 $ for $N \to \infty$ for all $q$. The weak dependence on the system size observed for $\delta =6$ suggests that $\delta^*$ is close to $6$. In fact, we can obtain a rough estimate of $\delta^*$ by considering the dependence of $\langle s_c \rangle  $ on $\delta$ for different system sizes, as shown in Fig.\ \ref{fig:11} for   $q=10$.  The critical value $\delta_c \approx 5.9$ was estimated by the intersection of  the curves that fit the data of  $N=2^{15}$ and  $N=2^{16}$. In addition, we find that  $\langle s_c \rangle \approx 0.5$ at $\delta = \delta_c$.  A similar analysis for $q=20$ and $q=30$ yielded the same estimate for $\delta_c$ so we conjecture that
$\delta^* \approx 6$.  
  
 %-----------------------------------------------------
%-----------------------------------------------------

The result that $\langle s_c \rangle \to 0 $ for $\delta > \delta^*$ regardless of the value of $q$ holds  true provided that $q$ is kept fixed while the system size $N$ approaches the thermodynamic limit. However, in the case $N$ is fixed and  $q$ becomes arbitrarily large then the agents are likely to be moving all the time (without interacting as the probability of finding an agent with a common feature is very low) and  the topology of the influence network becomes a sequence of RGGs. This scenario is discussed  in  \ref{App_q}.

%-----------------------------------------------------
%-----------------------------------------------------

\begin{figure}
\centering  
\includegraphics[width=0.48\textwidth]{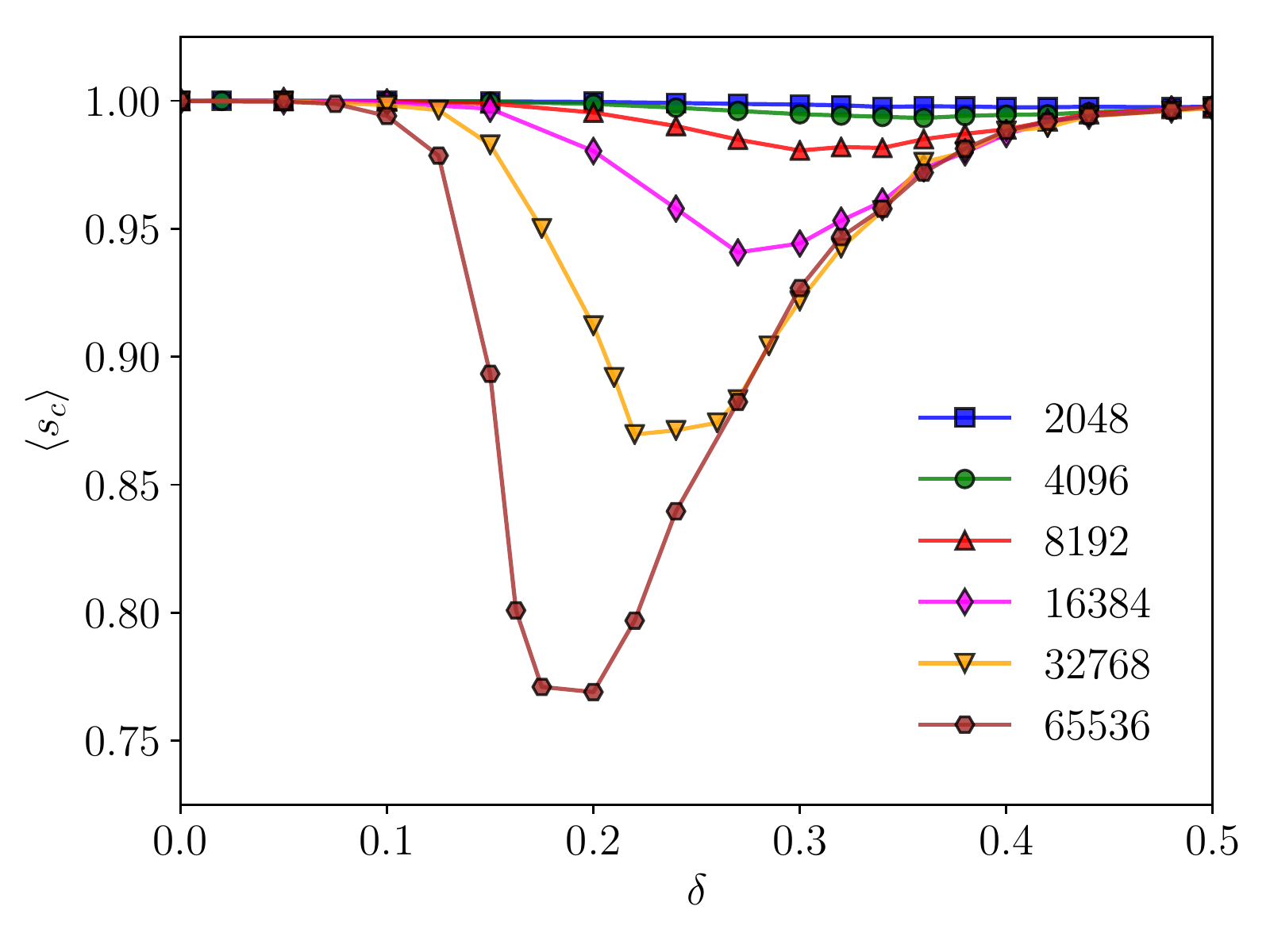} 
\caption{Mean fraction of agents in the largest component  $\langle s_c \rangle $ as function of the step size $\delta$  for $q=10$ and systems of size $N= 2^l$ with $ l=11, \ldots, 16$,  as indicated. 
 }  
\label{fig:12}  
\end{figure}

 Actually, the  effect of the step size $\delta$ on the connectedness of the influence network is way more complicated than that suggested in Fig.\ \ref{fig:11}. For instance, the results of Fig.\ \ref{fig:4} have already showed that for $q \in \left  [ 4,14 \right ]$ and $N = 2^{16}$  the network is more fragmented for $\delta=0.2$ than for $\delta=0.4$, which contrasts with the monotonous decrease of $\langle s_c \rangle $ with increasing $\delta$ exhibited in Fig.\ \ref{fig:11}. 
 This situation is seen more clearly in Fig.\ \ref{fig:12} that exhibits 
 the    region of small $\delta$ for  different system sizes and reveals  the existence of a  valley in the curve $\langle s_c \rangle $ vs.\ $\delta$  that becomes deeper and closer to $\delta=0$ as $N$ increases. (We recall that $\langle s_c \rangle =1$ for $\delta =0$ in the thermodynamic limit.)  The fragmentation observed for small $\delta$ is reminiscent of the fragmentation observed for small $q$ in Fig.\ \ref{fig:5} and, as in that case, we can only offer  a conjecture  about the asymptotic behavior of $\langle s_c \rangle$. Accordingly, we speculate that $\langle s_c \rangle$ vanishes continuously at $\delta_c$ for some  $\delta_c > 0$ so that  in the thermodynamic limit $\langle s_c \rangle$ jumps from $1$ to $0$ as  $\delta$ departs from $0$. We find it hopeless to estimate the continuous transition point $\delta_c$, which may be arbitrarily close to $0$,  because  $\langle s_c \rangle$ decreases very slowly with $N$ (see Fig.\ \ref{fig:12}), thus making  the  simulation of  the model near the continuous transition point   computationally impracticable.  
 
%-----------------------------------------------------
%-----------------------------------------------------
 
\begin{figure}
\centering  
 \subfigure{\includegraphics[width=0.48\textwidth]{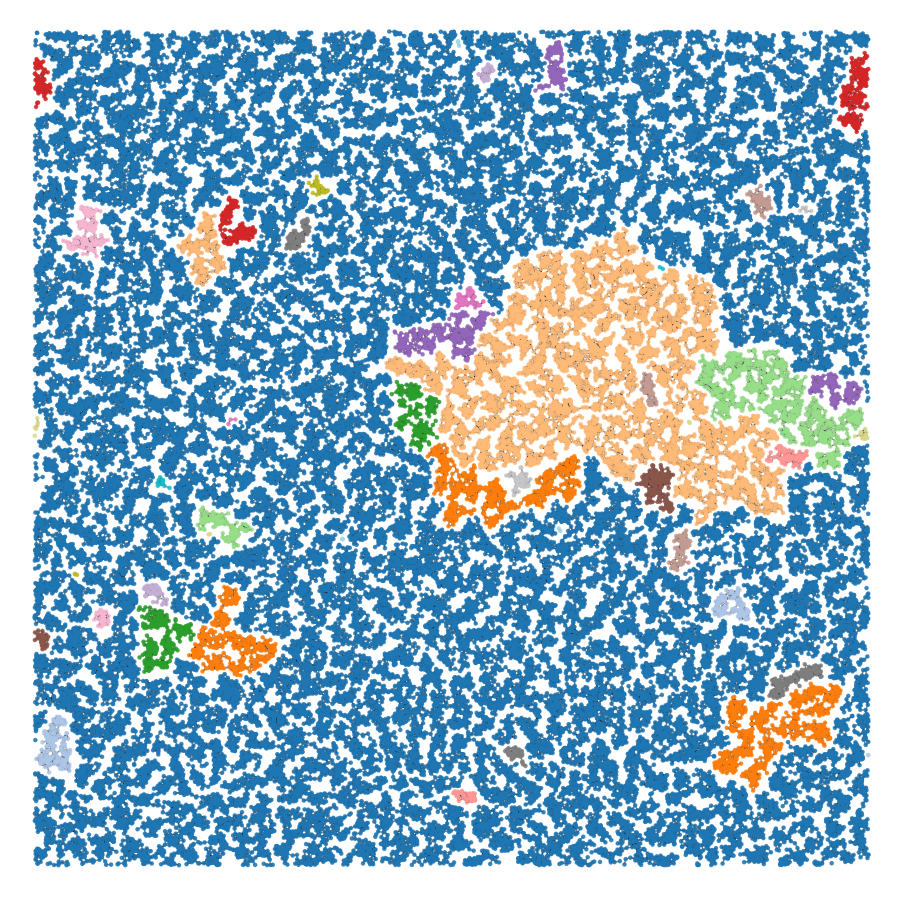}} 
 \subfigure{\includegraphics[width=0.48\textwidth]{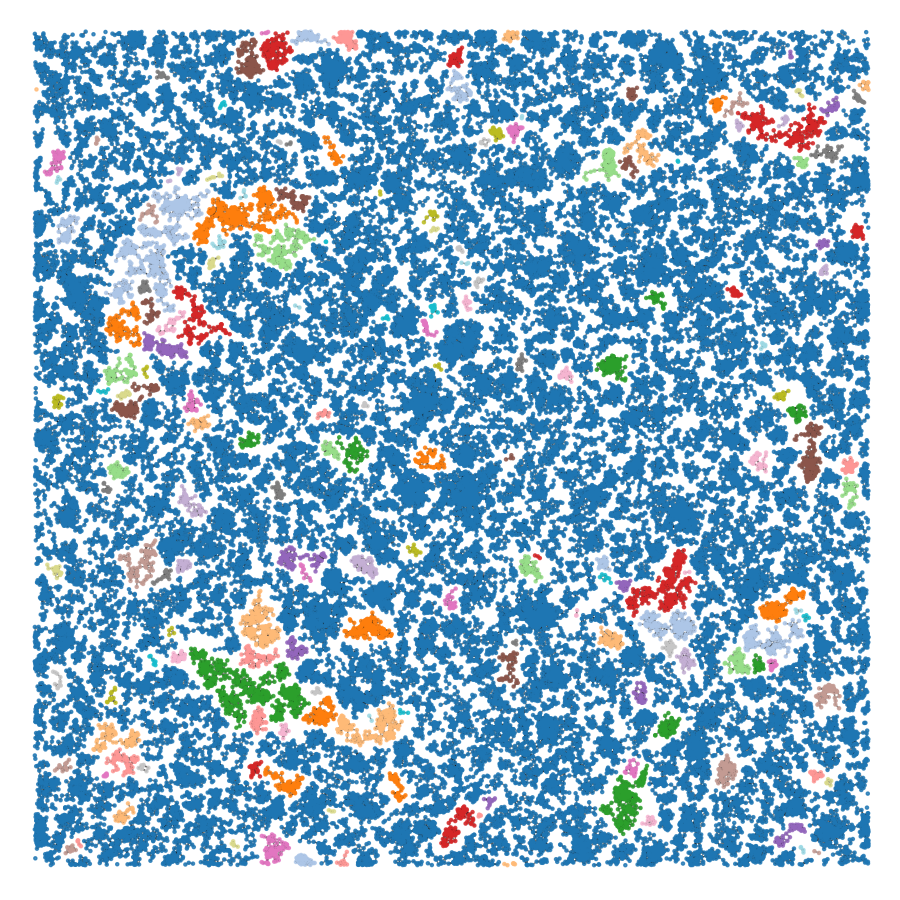}}
\caption{Snapshots of two absorbing configurations showing the components of the influence network for $N=2^{16}$, initial cultural diversity $q = 10$, step sizes  $\delta=0.2$ (upper panel) and
$\delta=5.6$ (lower panel). For  $\delta=0.2$ we have $s_c = 0.77$ and $n_c = 0.0007$, whereas for   $\delta=5.6$,  $s_c = 0.78$ and $n_c = 0.003$.
 }  
\label{fig:13}  
\end{figure}

 In order to gain an insight into the nature of
 the mildly fragmented regimes for small and large step sizes, we present in Fig.\ \ref{fig:13} snapshots of two absorbing configurations for $\delta =0.2$ and 
 $\delta = 5.6$, which exhibit largest components with approximately the same  size (viz., $s_c \approx 0.8$). It is evident that what distinguishes these configurations is the presence of other large components for small $\delta$, which could also be inferred by the  fact that, other things being equal,  there are many more components in the configuration  for $\delta =5.6$ than in the configuration for $\delta =0.2$.  
 
 %-----------------------------------------------------
 
Before concluding, a word is in order about the microscopic mechanism that leads to the severe fragmentation regime for small $\delta$. This unexpected phenomenon can be understood with aid of the pair correlation function $g_\delta \left ( r \right )$ shown in Fig.\ \ref{fig:B2} of \ref{App_top} and of  the absorbing configuration  shown in the upper panel of Fig.\ \ref{fig:13} for $\delta = 0.2$. The small step size  enhances the effect of social influence since 
$\delta \ll \alpha$  guarantees that many  interactions between a same pair of agents will happen before one of them has the  chance to leave  the influence neighborhood of the other, which results in a relatively high density of pairs separated by  distances smaller  than $r=\alpha$.  However, since for  small step sizes  the movement is practically continuous,  the agents fine tune their motions to create exclusion regions where they would be uncomfortable, producing a dip in the pair correlation function for $r  \in \left [ 2 \alpha, 3 \alpha \right ] $ which explains the extraordinary  similarity between the areas and shapes of the empty and occupied regions in the upper panel of Fig.\ \ref{fig:13}. Of course,  exclusion regions with linear sizes greater than $\alpha$ lead to the fragmentation of the influence network.

\begin{figure}
\centering  
 \subfigure{\includegraphics[width=0.48\textwidth]{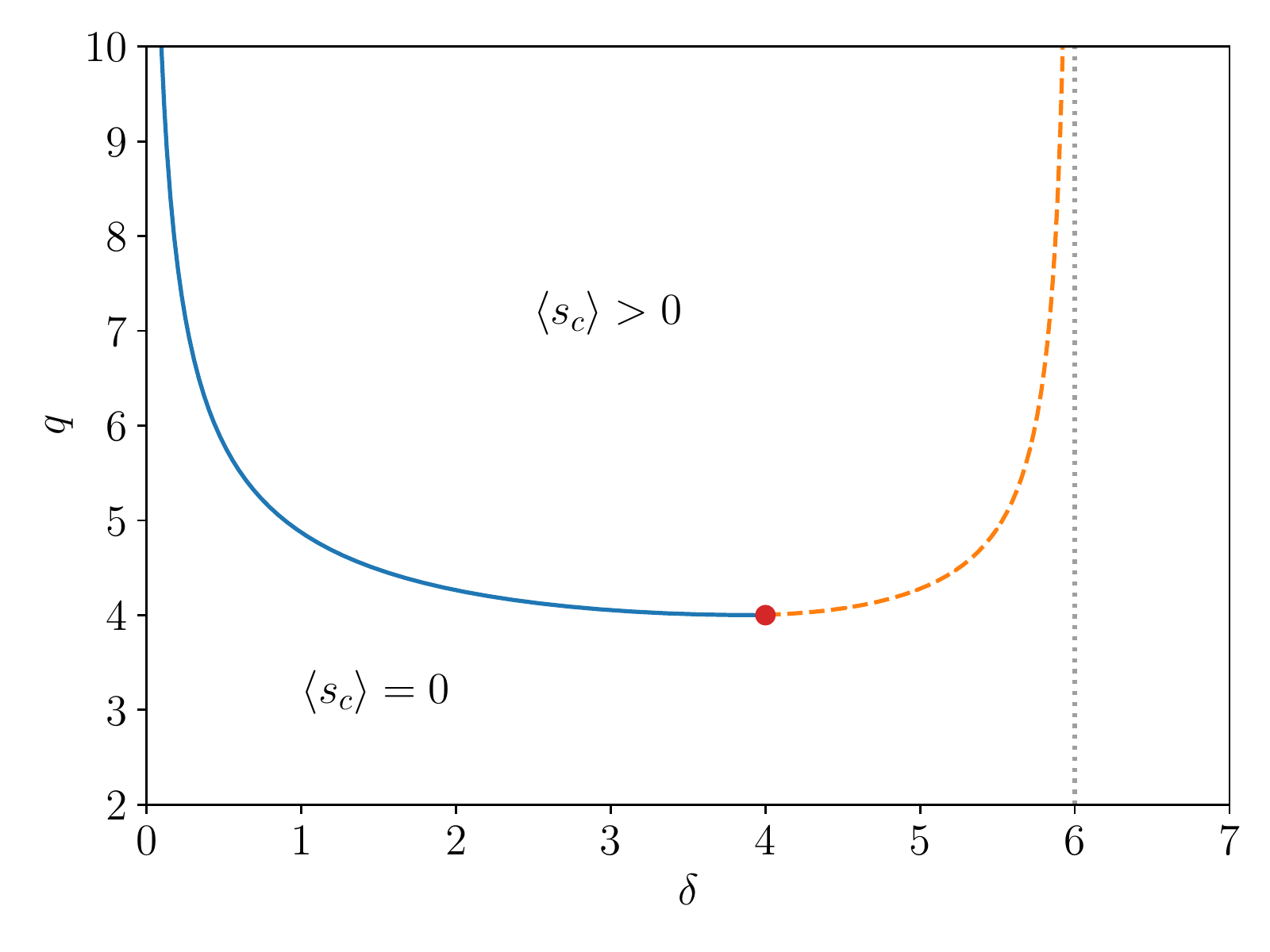}} 
 \subfigure{\includegraphics[width=0.48\textwidth]{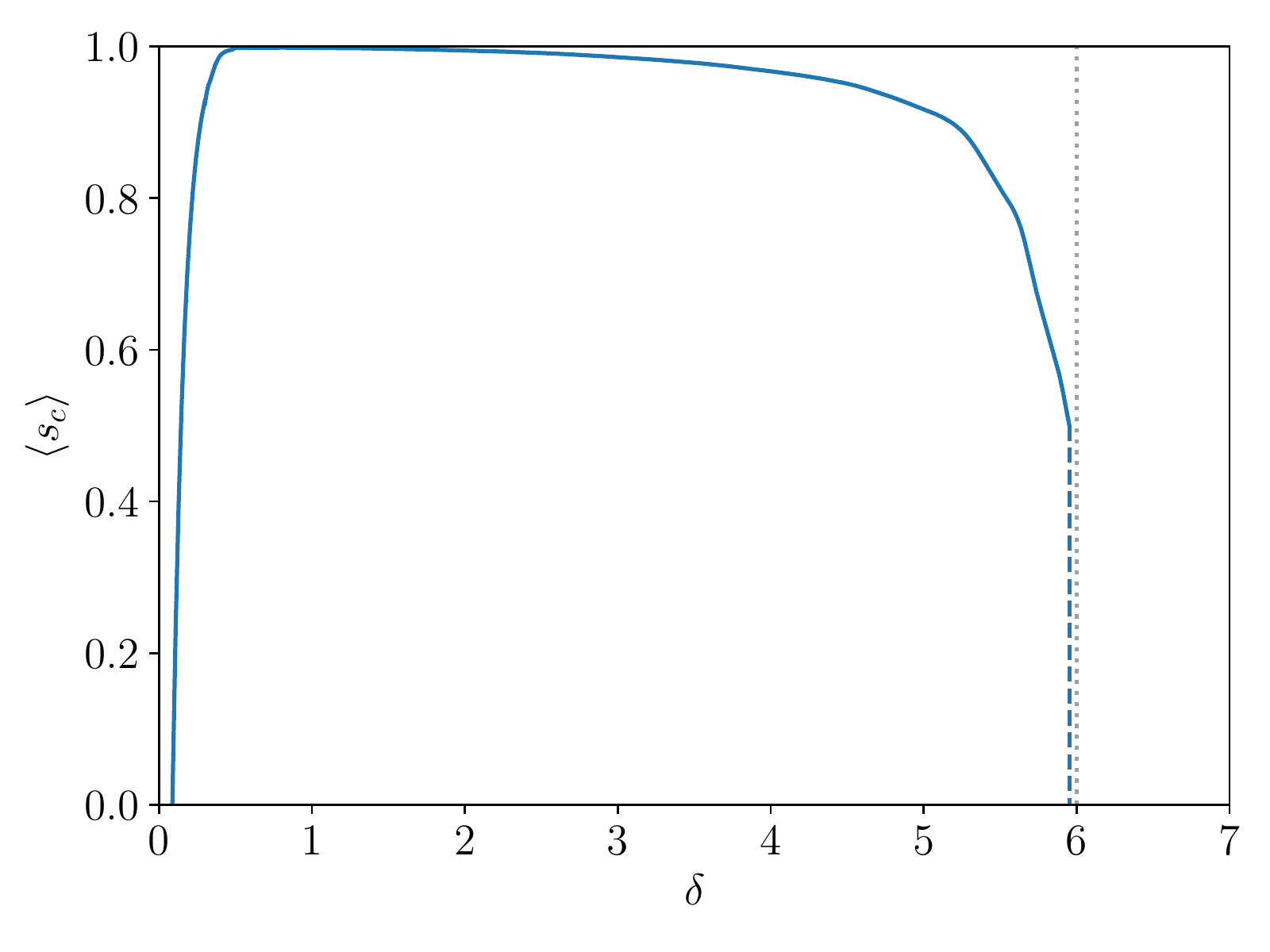}}
\caption{Schematic phase diagram (upper panel) showing  the continuous (solid curve) and discontinuous (dashed curve)  transition lines that separate the severely ($\langle s_c \rangle =0$) and mildly ($\langle s_c \rangle > 0$) fragmented regimes. The lower panel shows a schematic  representation of the order parameter in the thermodynamic limit for $q = 10$.
 }  
\label{fig:14}  
\end{figure}

Finally, to summarize our main results we offer in Fig.\ \ref{fig:14} a schematic portrayal of the phase diagram in the space  $(\delta,q)$ showing the continuous and the discontinuous transition lines. We recall that for $\delta > \delta^* \approx 6$ the absorbing configurations are severely fragmented regardless of the value of $q$.  This figure shows also a schematic depiction of the dependence of the order parameter $\langle s_c \rangle$ on $\delta$  in the thermodynamic limit,  which was produced by the extrapolations of the  finite size results exhibited in  Figs.\ \ref{fig:11} and \ref{fig:12} for $q=10$.

 %-----------------------------------------------------

%\clearpage
%\newpage

 \section{Discussion}\label{sec:disc}
 
Axelrod's model exhibits cultural diversity, in spite of the homogenizing effect of social influence,  thanks to the rule that excludes the interactions between individuals that differ from each other  in all their
cultural features \cite{Axelrod_97}.  Since in the original formulation of the model the agents are fixed at the nodes of a  network, this rule amounts to the existence of a substantial number of uncomfortable (i.e., culturally isolated) agents that are stock-still away from their likes (see Fig.\ \ref{fig:3}). A natural way to mitigate this frustration is to allow the agents to move away from  their cultural antagonists. Here we allow the agents to move a fixed distance $\delta$ (step size) in random directions in the plane with a probability that depends on the maximum value of the cultural similarity evaluated over the agents in their influence neighborhoods. In particular, the greater that cultural similarity, the more comfortable the agent is  and so the  lesser the probability that it moves.   We recall that in this paper we define the influence neighborhood of an agent as the region limited by  a circle of radius $\alpha = 1.75$ centered at the agent, and that the density of agents in the square box of linear size $L$  is set to $\rho=1$.

We hasten to note that an agent does not purposely move towards or away  other agents in our model. However,  the  rule of motion which prescribes that the agent will stay put if it has an  identical neighbor and that it will jump a distance $\delta$ in any direction if all its neighbors are  antagonists  results  in  an effective evasive behavior towards cultural antagonists, specially if $\delta > 2 \alpha$. The explicit spatial scenario we consider here allows a clearer visualization and interpretation of the self-organized network components (see Fig.\ \ref{fig:13} and Fig.\  \ref{fig:B2}) as compared with approaches based on the rewiring of links \cite{Min_17,Murase_19}. In addition, our model offers a necessary development of the static scenario introduced in  the original  formulation of Axelrod's model since spatial proximity and mobility are key elements  of the celebrated propinquity effect of social psychology, which  is the tendency for people to form social bounds  with those whom they encounter often  \cite{Festinger_63}.
 The drawback of our approach is that the simulation times to reach the absorbing configurations are much longer than for the  original model,  mainly due to the additional  rules of motion and the need to keep a record of the positions of the agents  in order to determine their influence neighborhoods.  These  add-ons limited the maximum system size we could  simulate in a feasible time to  $N =2^{18}$, although it is clear from  Fig.\ \ref{fig:12} that much larger system sizes are necessary to bypass the strong  finite size effects observed for small $\delta$.

We find that the introduction of the comfort-driven mobility  in Axelrod's model produces  a variety of startling results. In particular, we find that the  influence network is  fragmented for low initial cultural diversity $q$. We argue that this fragmentation is extreme, in the sense that  all components of the influence network are of microscopic size  in the  thermodynamic limit (see Fig.\ \ref{fig:6}).
%Although in the case of  static agents  a low initial cultural diversity leads to  a  monocultural regime (see Fig.\ \ref{fig:2}), in the case of comfort-driven mobility  it results in extreme spatial fragmentation  with the  few distinct cultures coexisting inside the small components since  $\langle s_d \rangle/ \langle s_c \rangle \to 0 $  for large $N$.
 As  $q$ increases, the system transitions  to
a mildly fragmented regime, which is characterized by the presence of a macroscopic component (i.e., $\langle s_c \rangle > 0 $  for  $N \to \infty$). The value of the step size $\delta$ determines whether the  transition between the regimes of severe and mild fragmentation is continuous or discontinuous:  the fragmentation transition is continuous for $\delta < 4.2$ and discontinuous otherwise.   
%The nature of the transition is determined by the dependence of $\langle s_c \rangle$  on $N$ in the mildly fragmented regime, which, in turn,  determines whether  the curves  of $\langle s_c \rangle $ vs.\ $q$  cross or not for different system sizes.
 The discontinuous transition between the two fragmentation regimes disappears altogether for $\delta > 6$, so the absorbing configurations are severely fragmented for all values of the initial diversity $q$. This severe fragmentation is not surprising for large $\delta$, since  in this case an agent can quickly  survey the entire plane to find  its cultural likes and then freeze close to them. However,   we find that   the severe fragmentation occurs for small $\delta$ as well (see Fig.\ \ref{fig:4}),  thus  resulting  in a discontinuous jump of the relative size of the largest component at $\delta=0$  since the influence network is connected in the static limit. Regardless of the fragmentation regime, we find  that the relative size of the largest cultural domain $\langle s_d \rangle$ vanishes in the thermodynamic limit, so the absorbing configurations are always multicultural.

The feedback between mobility  and  cultural similarity is responsible for these nonintuitive results, which make  the behavior of Axelrod's model  with comfort-driven mobility nonobvious logical consequences  of the  interaction rules, attesting thus the value and the need of the simulation model \cite{Reijula_19}. In fact,  the agent-based model  proposed   by Axelrod in the late 1990s  
   has endured the  test of time so far, probably because it exhibits the right balance between simplicity and realism   as well as very intriguing  critical phenomena \cite{Castellano_00,Vilone_02,Peres_15}.  In this vein, introduction of comfort-driven mobility in Axelrod's model produced
results of interest for the social sciences, such as  the prevention of the formation of large cultural domains  and   the spatial  segregation of the agents,  as well as for statistical physics, such as the continuous and discontinuous fragmentation transitions of the influence network. Thus our results reaffirm Axelrod's model as a  paradigm for idealized models of collective behavior \cite{Goldstone_05}.

\bigskip

\acknowledgments
The research of JFF was  supported in part 
 by Grant No.\  2017/23288-0, Fun\-da\-\c{c}\~ao de Amparo \`a Pesquisa do Estado de S\~ao Paulo 
(FAPESP) and  by Grant No.\ 305058/2017-7, Conselho Nacional de Desenvolvimento 
Cient\'{\i}\-fi\-co e Tecnol\'ogico (CNPq).
SMR  was supported by grant  	15/17277-0, Fun\-da\-\c{c}\~ao de Amparo \`a Pesquisa do Estado de S\~ao Paulo 
(FAPESP). Research carried out using the computational resources of the Center for Mathematical Sciences Applied to Industry (CeMEAI) funded by FAPESP (grant 2013/07375-0).

\appendix

\section{Limit of high $q$}\label{App_q}

 %-----------------------------------------------------
\begin{figure}[b!]
\centering  
 \subfigure{\includegraphics[width=0.45\textwidth]{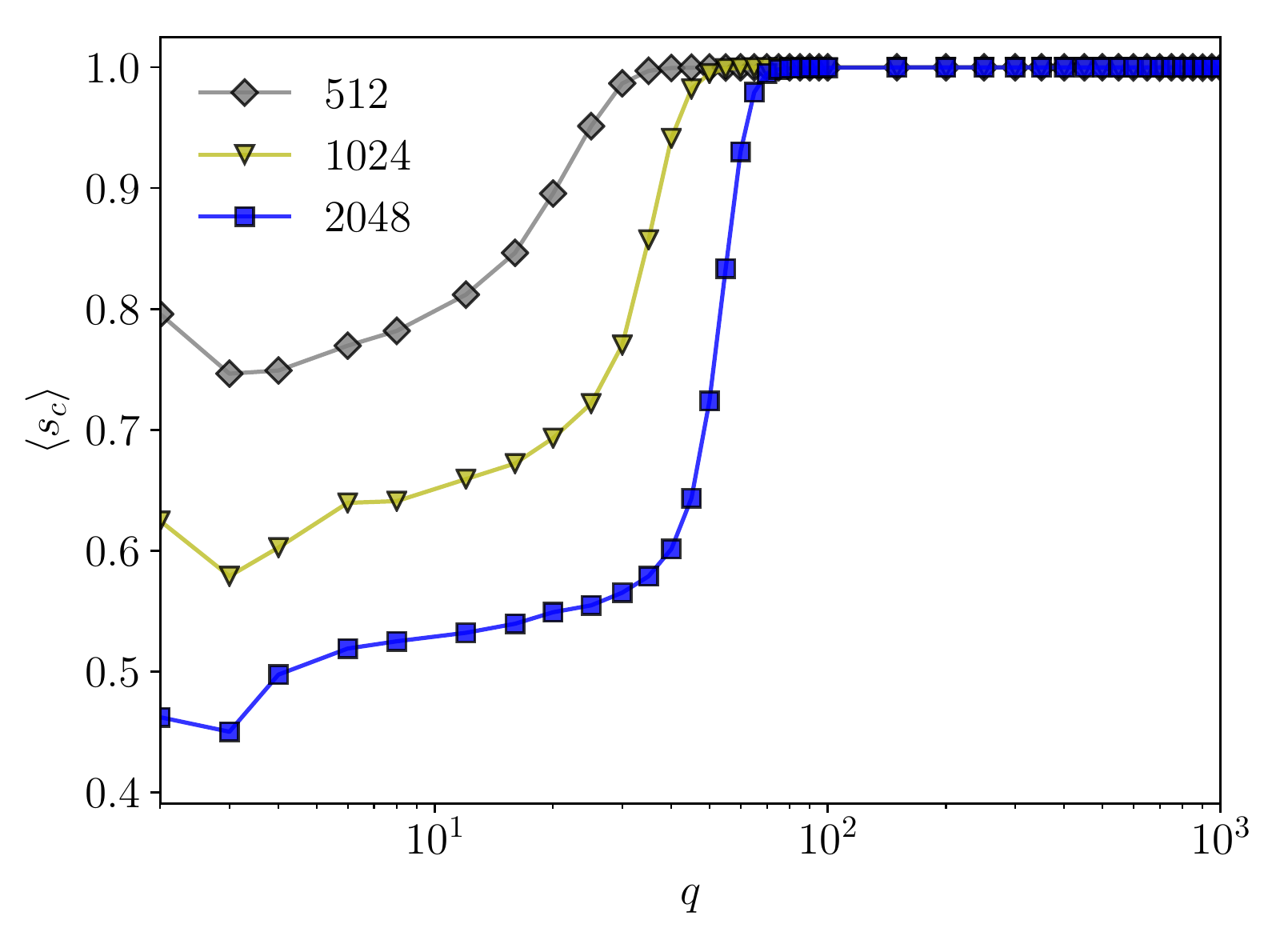}} 
 \subfigure{\includegraphics[width=0.45\textwidth]{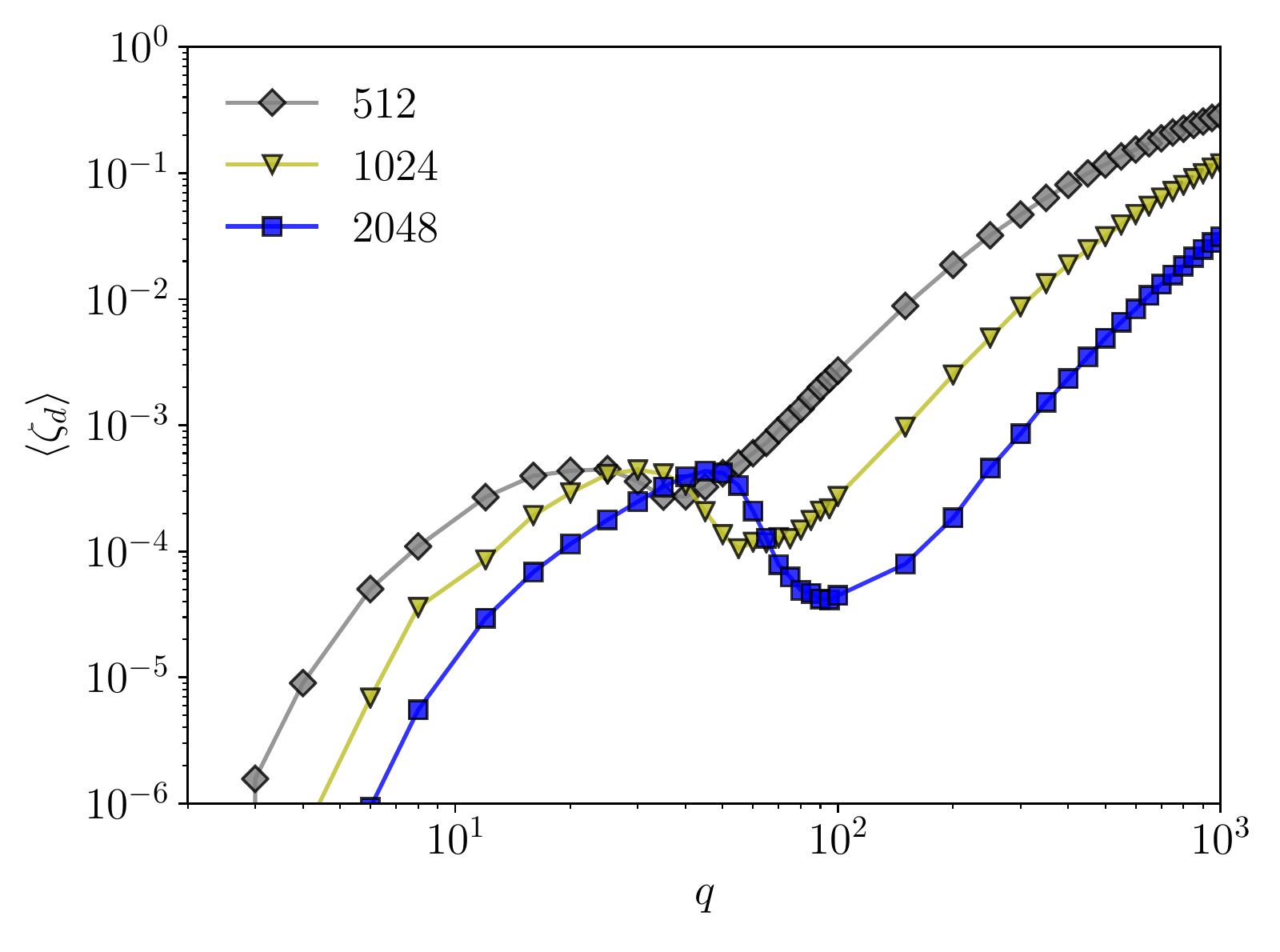}}
\caption{Mean fraction of agents in the largest component  $\langle s_c \rangle $ (upper panel) and  mean fraction of culturally isolated agents $\langle \zeta_d \rangle $ (lower panel) as functions of the  initial cultural diversity $q$ for $\delta=7$ and systems of size $N= 2^l$ with $ l=9, 10,11$,  as indicated. 
 }  
\label{fig:A1}  
\end{figure}
%-----------------------------------------------------

For very high $q$ and fixed $N$, we expect that the absorbing configurations will not fragment for  $\delta >0$ because the agents are likely to wander around the square box  without interacting, since the probability of finding an agent with a common feature is very low for $q \gg 1$.   The  topology of the influence network  becomes then a sequence of random geometric graphs (RGGs). To check this presumption  we present in Fig. \ref{fig:A1} the mean fraction of agents in the largest component  $\langle s_c \rangle $ as well as the fraction of  non-interacting (i.e., culturally isolated) agents  $\langle \zeta_d \rangle $ for much larger values of $q$ than those used in the main text and for relatively small system sizes $N$. We recall that, for the step size $\delta=7$ used to generate these results, the influence network is severely fragmented in the thermodynamic limit. It is clear from our results that, for fixed $N$, an increase of $q$ will  defragment the influence network according to our expectation. However, as $N$  increases, the defragmentation occurs for larger and larger values of $q$, which agrees with our conclusion that for any finite $q$ the influence network is severely fragmented in the thermodynamic limit.

The lower panel of  Fig. \ref{fig:A1}, which exhibits the fraction of non-interacting agents, confirms that for the range of $q$ and  $N$ considered in the main text, those agents are very rarely observed in the absorbing configurations produced by the comfort-driven mobility.  In fact, even for the small system sizes show in this figure, the presence of  the non-interacting agents is noteworthy for very large $q$ only. More explicitly, since for $\delta >0$  an  agent can, in principle,  enter the influence neighborhood of any other agent in the square box, the probability that a particular agent becomes culturally isolated  in the initial setup of the system is $\left ( 1 - 1/q \right )^{NF} \approx \exp \left ( -NF/q \right)$ so that $q$ must be on the order of $NF$ to guarantee that $\langle \zeta_d \rangle $ is on the order of 1.

%The non-monotonic dependence of $\langle \zeta_d \rangle $ on $q$ indicates that this quantity is influenced by the connectedness of the influence network as well. Somewhat surprisingly, a network that exhibits intermediate size   has fewer non-interacting agents than  a network  composed of small size components only.

\section{Topology of the absorbing configurations}\label{App_top}

\begin{figure}[t!]
\centering  
\subfigure{\includegraphics[width=0.45\textwidth]{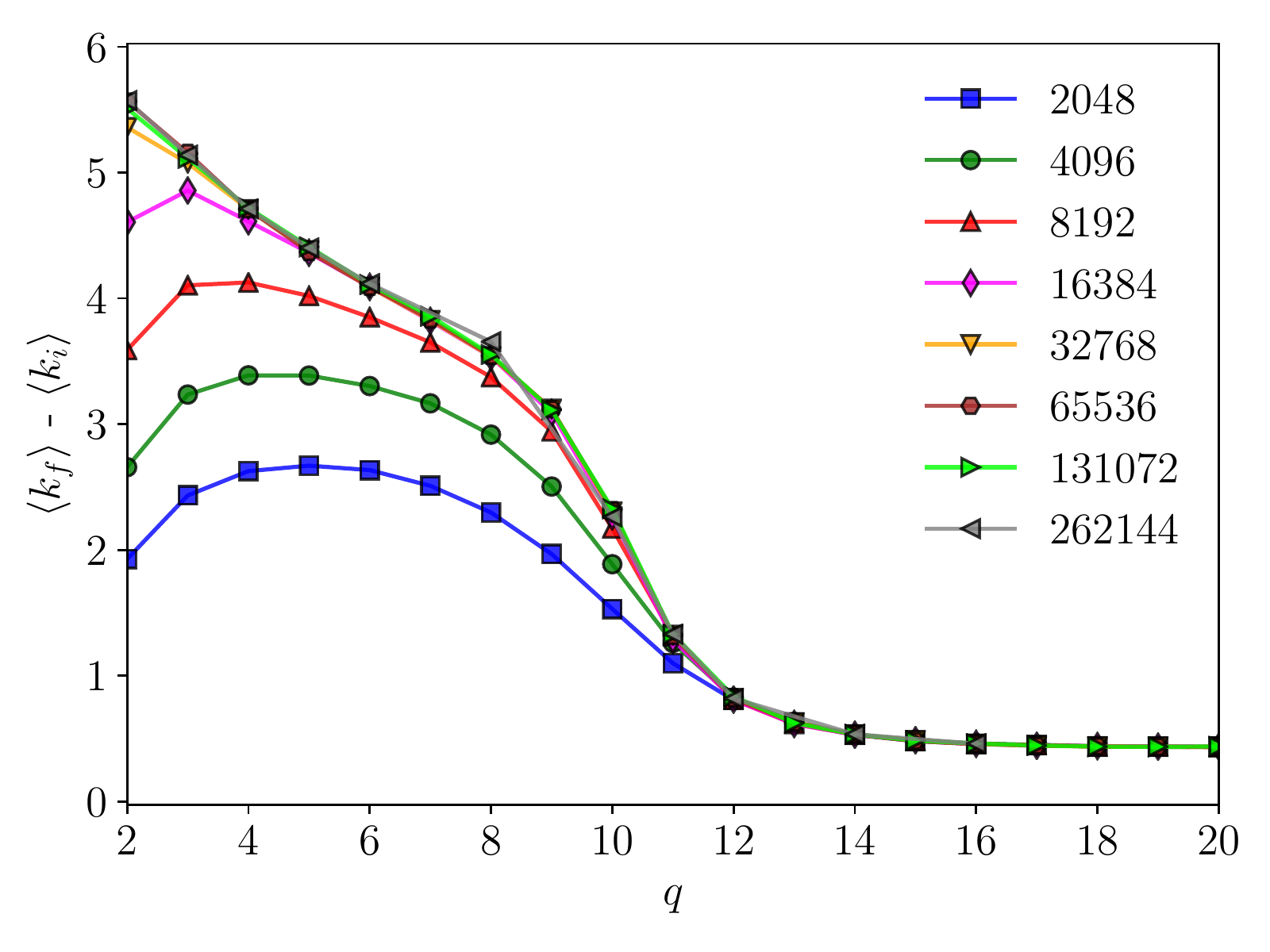}} 
 \subfigure{\includegraphics[width=0.45\textwidth]{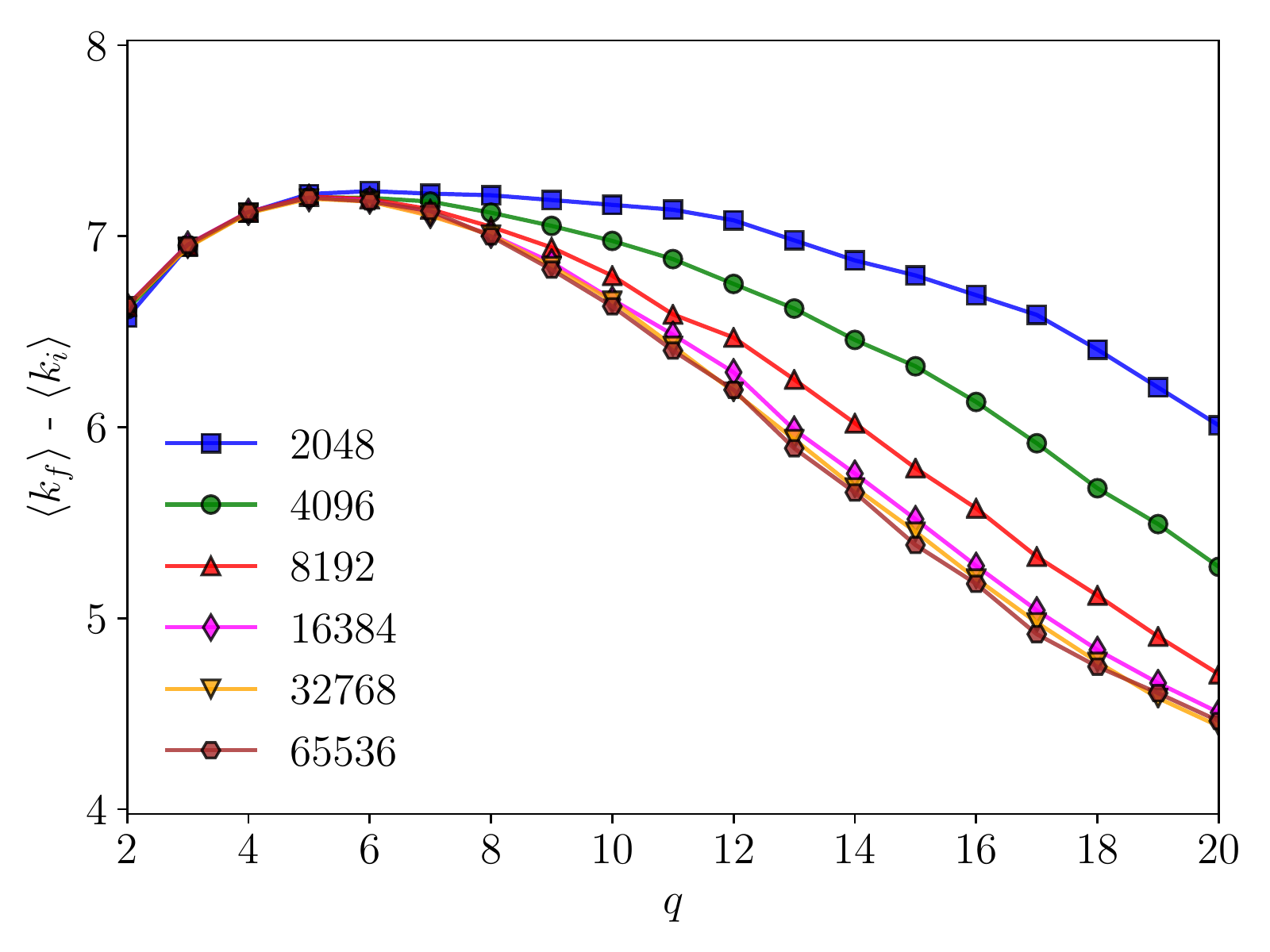}}
\caption{Difference between the final and initial mean degree of the influence networks  as a function of the  initial cultural diversity $q$ for $\delta = 0.4$ (upper panel)   and $\delta=5$  (lower panel)  and systems of size $N$ as indicated. 
 }  
\label{fig:B1}  
\end{figure}

The initial topology of the influence network is a random geometric graph (RGG), but the comfort-driven mobility leads to absorbing configurations that are characterized, in principle, by  different topologies.  Here we offer a brief characterization of those topologies focusing on the mean degree (i.e.,  the mean number of links an agent  has to other agents) of the influence network, which is the simplest and most impacting  property of the topology.  Figure \ref{fig:B1} shows the difference between the mean degree of the absorbing and the initial configurations for $\delta =0.4$ and $\delta = 5$ (see Figs.\ \ref{fig:6} and \ref{fig:8} for the size of the largest component corresponding to these step size values).
We recall that for $\alpha =1.75$ and $\rho = 1$, the mean degree of the initial RGGs  is $ \langle k_i \rangle \approx 9.62 $.

%-----------------------------------------------------
%-----------------------------------------------------

%-----------------------------------------------------
\begin{figure}[b!]
\centering  
\subfigure{\includegraphics[width=0.45\textwidth]{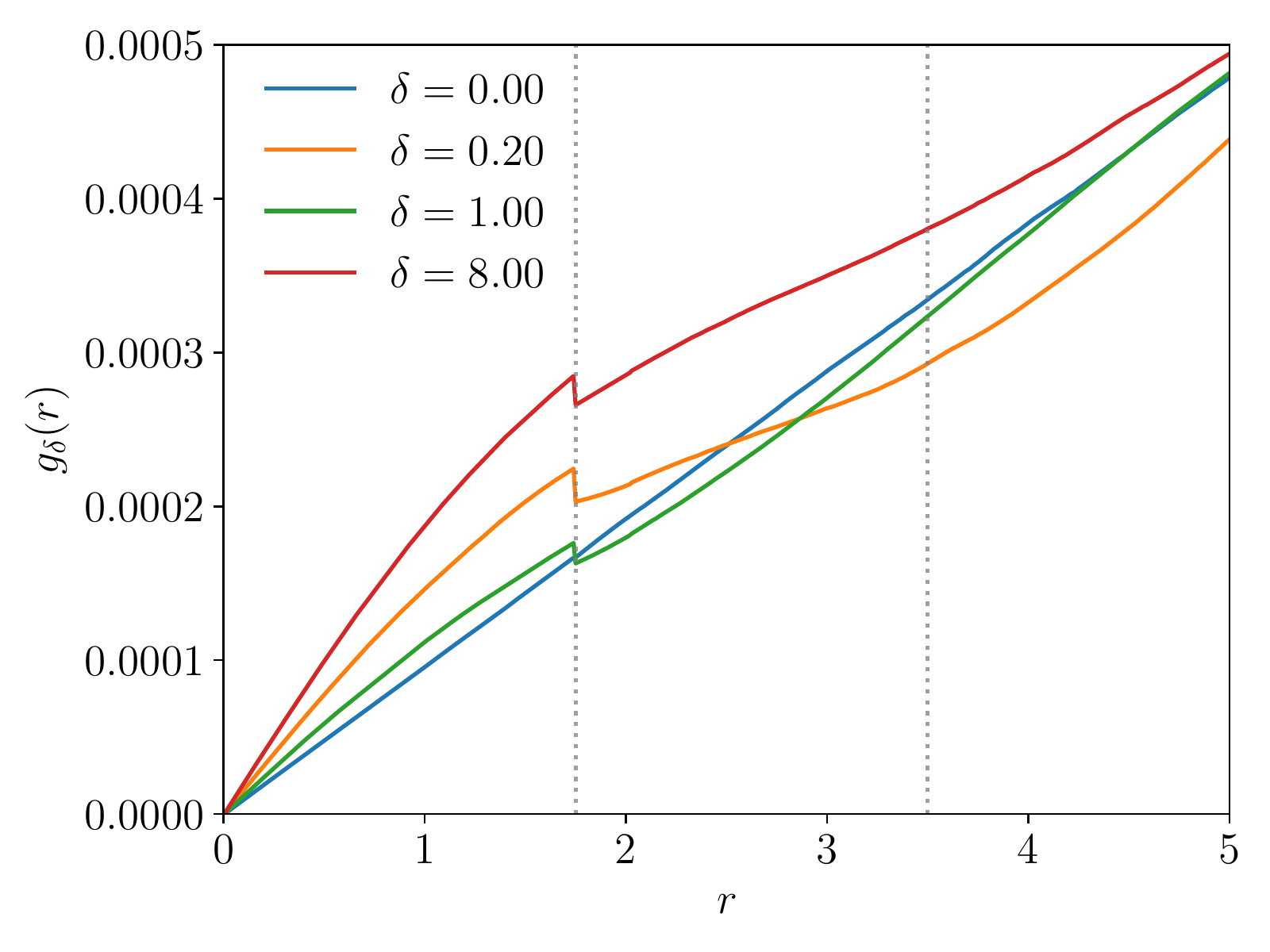}} 
 \subfigure{\includegraphics[width=0.45\textwidth]{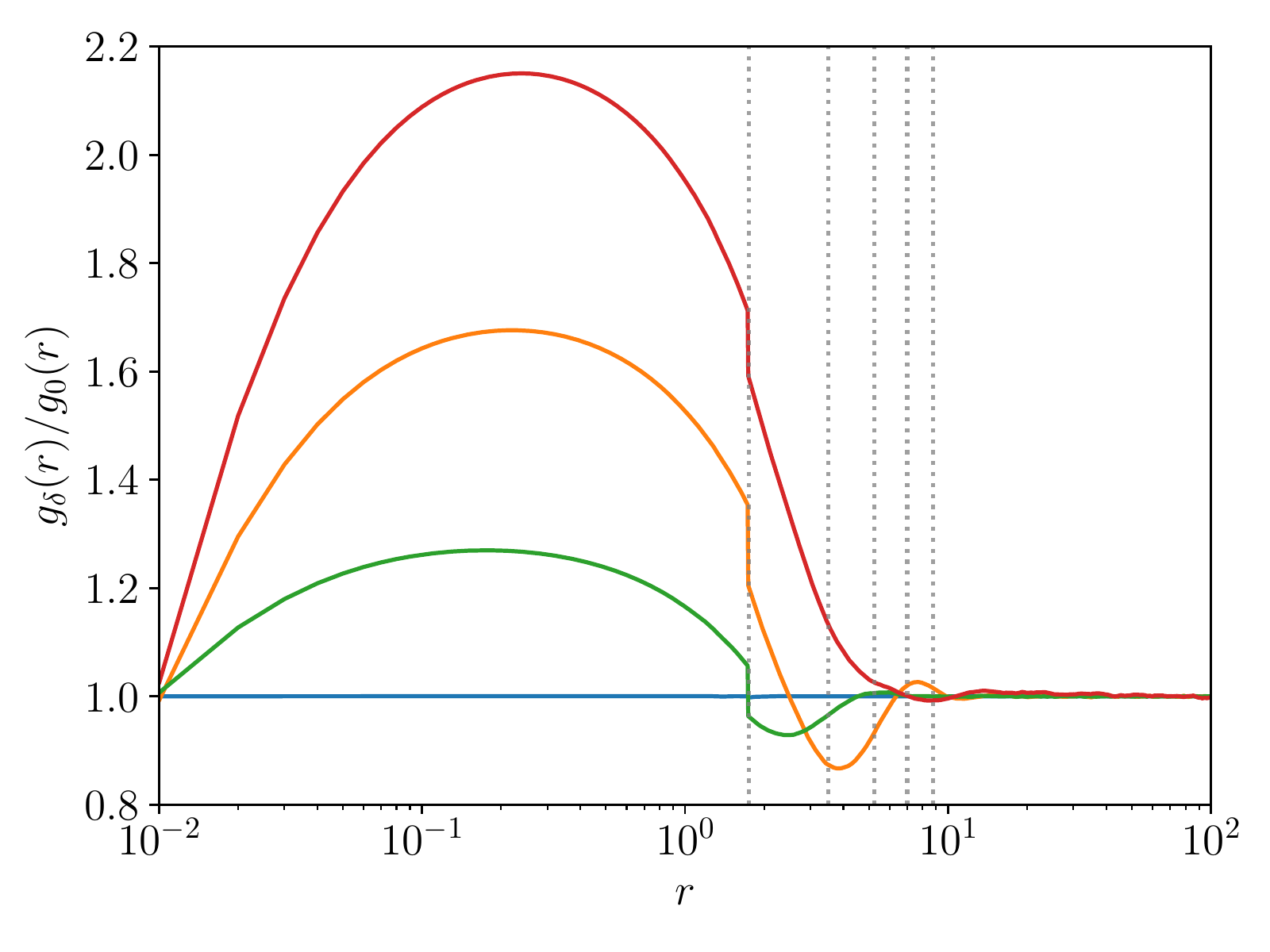}}
\caption{Pair correlation function $g_\delta$  (lower panel)) for the comfort-driven mobility with step sizes $\delta = 0, 0.2, 1, 8$, as indicated.    The upper panel shows the  ratio $g_\delta/g_0$ where $g_0 \left ( r \right ) \approx 2 \pi r/N$  is the pair correlation function for the RGGs ($\delta=0$). The system size is $N=2^{16}$ and the initial cultural diversity is $q=10$. The vertical dotted lines are $r=\alpha, 2 \alpha, \ldots, 5 \alpha$ with $\alpha=1.75$.
 }  
\label{fig:B2}  
\end{figure}
%-----------------------------------------------------

A first point to be noted in Fig.\ \ref{fig:B1} is that the mean degree tends smoothly  to its asymptotic value  in the thermodynamic limit, so the fragmentation phase transitions have no consequences on the first moment of the degree distribution.  For $\delta =0.4$ (upper panel), we note that  the absorbing configurations of the severely fragmented regime  exhibit a significantly  larger degree than those of the mildly  fragmented regime, which have a mean degree very close to that of the RGGs.  This result implies a high density of agents within the components and, in fact, there is a high correlation between the  mean degree and the density of agents within the largest component (see lower panels of  Figs.\ \ref{fig:6} and \ref{fig:8}). It is curious that the mean degree increases with increasing $N$ for $\delta =0.4$, but decreases with $N$ for $\delta=5$. Moreover, the regions with the  strongest  finite size effects are  reversed for these values of $\delta$.  For fixed $q$, the mean degree of the influence network  increases with increasing  $\delta$,  which means that one should expect the  clusters of agents to be more compact for large $\delta$, in agreement with  the spatial organization depicted in  the snapshots of  the absorbing configurations of  Fig.\ \ref{fig:13}.

The fragmentation of the influence network  in components  implies that the density of agents is not homogeneous along the
square box (see Fig.\ \ref{fig:13}). We recall that  the  density of agents is homogeneous in the initial configurations for all $\delta$ as well as in the absorbing configurations for $\delta=0$, since the influence networks are RGGs in those cases.  In Fig.\ \ref{fig:B2} we  quantify the spatial distribution of the agents  by calculating the fraction of pairs of agents whose  distance is between $r$ and $r+\delta r$, which we  denote by $g_\delta \left ( r \right ) \delta r$. Clearly,  $g_\delta \left ( r \right )$  is the usual pair correlation function of Statistical Mechanics. In this figure we use $\delta r = 0.01$ for a system of  $N=2^{16}$ agents, so the length of the sides of the square box is $L=256 \gg \delta r$. In addition, for $q=10$ we pick representative values of the step size, viz. $\delta=0.2, 1, 8$, that lead to the different fragmentation scenarios revealed in Figs.\ \ref{fig:11} and \ref{fig:12}.  For the RGGs of the static case we have $g_0 \left ( r \right ) = 2 \pi r/N$ for not too large $r$. Most interestingly, $g_\delta \left ( r \right )$ exhibits a discontinuity at $r = \alpha$ for $\delta > 0$.  Since the absorbing configurations produced by the comfort-driven mobility  exhibit  a greater number of agents separated by distances $r < \alpha$  compared to the RGGs, this  type of mobility results in an effective attraction between culturally similar agents.  In the case of $\delta = 0.2$,  which corresponds to  fragmented  networks, we observe a high density of pairs of agents separated by  distances  less  than  $r=\alpha$ and a low density of pairs separated by distances  around  $r= 2 \alpha$. Of course, the pair densities are high or low depending on whether they are greater or less than the pair density of  the RGG. The two absorbing configurations depicted in Fig.\ \ref{fig:13}  illustrate quite nicely the results of  Fig.\ \ref{fig:B2} since we can see that the clusters are denser and larger for $\delta = 5.6$ than for $\delta = 0.2$. Moreover,  for $\delta=0.2$ the rarified regions, corresponding to the  dip at $r \approx 2 \alpha$ of the normalized pair correlation function, are as conspicuous as  the clusters,  corresponding to the peak at  $r \approx 0.2 $ (see lower panel of Fig.\ \ref{fig:B2}).

\end{document}